\documentclass{nature}

    \usepackage{authblk}
    \usepackage{amsmath,amssymb,amsfonts}
    \usepackage{mathptmx} 
    
    \usepackage{siunitx}
    \usepackage{svg}
    
    \usepackage{graphicx} 
    \usepackage{xcolor}
    \usepackage[font=small,labelfont=bf]{caption}
    \usepackage{float}
    \usepackage[version=4]{mhchem}
    \usepackage[section]{placeins}

    \usepackage[resetlabels]{multibib}

    \usepackage{fancyhdr}
    \fancypagestyle{plain}{} {
    }

    \newcites{sicite}{\_}

\newif\ifsubmit
\submitfalse

\newcommand{\eg }{\textit{e.g.}}

\ifsubmit
    \newcommand{\cat}[1]{}
    \newcommand{\johnpaul}[1]{}
    \newcommand{\can}[1]{}
    \newcommand{\todo}[1]{}
    \newcommand{\tocite}[1]{}

\else
    \definecolor{comments}{rgb}{0.1, 0.66, 0.1}
    \newcommand{\cat}[1]{[{\color{comments}CG: #1}]}
    \newcommand{\johnpaul}[1]{[{\color{comments}JPS: #1}]}
    \newcommand{\can}[1]{[{\color{comments}CL: #1}]}
    \newcommand{\todo}[1]{[{\color{red}TODO: #1}]}
    \newcommand{\tocite}[1]{[{\color{red}citation-#1}]}

\fi

    \PassOptionsToPackage{hyphens}{url}\usepackage{hyperref}

    \title{Analog content addressable memories with memristors}
    
    \author[1,*]{Can Li}
    \author[1,*]{Catherine E. Graves}
    \author[1]{Xia Sheng}
    \author[2]{Darrin Miller}
    \author[2]{Martin Foltin}
    \author[1]{Giacomo Pedretti}
    \author[1,*]{John Paul Strachan}
    
    \affil[1]{Hewlett Packard Labs, Hewlett Packard Enterprise, Palo Alto, CA 94304, USA}
    \affil[2]{Silicon Design Lab, Hewlett Packard Enterprise, Fort Collins, CO 80528, USA}
    \affil[*]{Emails: can.li@hpe.com; catherine.graves@hpe.com;
    john-paul.strachan@hpe.com }
    
    \date{}

    \makeatletter
    \let\newtitle\@title
    \let\newauthor\@author
    \makeatother
    
\begin{document}
    \maketitle
    \pagestyle{fancy}

    \begin{abstract}
    \begin{center} Abstract \end{center}
        A content-addressable-memory compares an input search word against all rows of stored words in an array in a highly parallel manner. While supplying a very powerful functionality for many applications in pattern matching and search, it suffers from large area, cost and power consumption, limiting its use. Past improvements have been realized by using memristors to replace the static-random-access-memory cell in conventional designs, but employ similar schemes based only on binary or ternary states for storage and search. 
        We propose a new analog content-addressable-memory concept and circuit to overcome these limitations by utilizing the analog conductance tunability of memristors. Our analog content-addressable-memory stores data within the programmable conductance and can take as input either analog or digital search values. Experimental demonstrations, scaled simulations and analysis show that our analog content-addressable-memory can reduce area and power consumption, which enables the acceleration of existing applications, but also new computing application areas.

    \end{abstract}
    \pagebreak

    \section*{Introduction}

    To increase power efficiency and cost performance, there is growing interest in computing architectures that allow for in-memory processing\cite{williams2017moore} in order to reduce data movement and address the memory wall. 
    In this vein, recent work has shown the promise of using non-volatile memory devices, or memristors, for accelerating matrix multiplication directly in memory arrays, accelerating a range of applications such as machine learning\cite{hu2018dpe,li2018traininig,ambrogio2018equivalent,bayat2018mlp,wu2017face}, analog signal processing\cite{li2018analog,sheridan2017sparse}, and scientific computing\cite{zidan2018pde,ibm2018mixed,sun2019pnas}.
    The performance improvements from this approach originate from two principles. First, computation is performed where the data is stored, removing the expensive power and latency costs of data movement between separate computing and memory units in a von-Neumann machine.
    Second, computation is performed in the analog domain, which provides exponential efficiency gains over digital, particularly at lower precision requirements. Each device performs analog computations that would otherwise require multiple digital elements. 
    Despite the great promise of this approach, demonstrations have thus far been limited to the acceleration of matrix multiplication via crossbars.

    Meanwhile, in-memory computational approaches in the digital domain have been extensively explored over the years\cite{zhang2015in_memory}.
    While many proposed circuit typologies have not been implemented in commercial systems, content addressable memory (CAM) and the related ternary CAM (TCAM) have stood as a notable exception\cite{pagiamtzis2006tutorial, meiners2010fast}.
    CAM/TCAM circuits natively perform a matching operation between an input data word (search key) and a stored set of data patterns in the CAM/TCAM array. The operation is highly parallel and another example of an in-memory operation, leading to extremely high throughput compare operations at low latency, and therefore commercial success in applications such as network routing\cite{chao2002router, mcauley1993router}, real-time network traffic monitoring\cite{xu2016survey}, and access control lists (ACL)\cite{bechtolsheim2002acl}.
    While powerful, CAM performance benefits come at the cost of large power and low memory density, limiting modern usage to high cost niche areas that demand high performance.
    Recent work has shown that utilizing non-volatile memristors (or resistive memory devices) in TCAM circuits reduces area and power\cite{chang2014vlsi, chang2015isscc,chang2016isscc,ibm2013_2t2r,graves2018regex,grossi2018leti, yiranchen2018tcam,li2016nvsim,yang2019NEtcam} and provides the flexibility to accelerate powerful finite state machines, particularly for Regular Expression matching used in Network Intrusion Detection Systems\cite{graves2018regex, graves2018icrc}. 
    However, nearly all memristor-based CAM designs utilize schemes similar to conventional static random-access-memory (SRAM) designs where the memristor only encodes binary states. The highly tunable analog conductance in memristor devices, with many stable intermediate states are not leveraged\cite{sheng2019aem}.
    An analog CAM design was proposed more than a decade ago that matches an input voltage with precise values stored in analog storage cells\cite{blyth2006analog_patent}, but has not been implemented likely due to practical concerns of high power and area as it requires large numbers of active comparators and inefficient array implementations.

    Here, we propose a memristor-based analog CAM that significantly increases data density and reduces operational energy and area for these in-memory processing circuits. 
    Our analog CAM design stores a range of values in each cell using the tunable conductance of memristive devices, and compares an analog input with this stored range to determine a match or mismatch. 
    The concept has been validated with proof-of-concept experiments, as well as simulations to establish performance and scalability. 
    When used to store narrow ranges as discrete levels, our analog CAM can be a direct replacement of digital CAMs while providing higher memory densities and smaller power consumption. 
    This may enable the use of CAMs for more generic scenarios\cite{batcher1974staran,tracy2016rf,guo2011resistive,guo2013acdimm} such as for associative computing 
    that otherwise struggle with the limited memory densities and high power consumption of conventional CAMs. 
    More importantly, our analog CAM can store wide intervals of  continuous levels, thereby enabling novel search and matching functionality in the analog domain.  
   The analog CAM cell presented here can also be searched with analog input signals, enabling the processing of analog sensor data without the need for an analog-to-digital conversion step.

    \section*{Results}
    \subsection{6-transistors 2-memristors analog content addressable memory} 
\begin{figure}
        \centering 
        \includegraphics{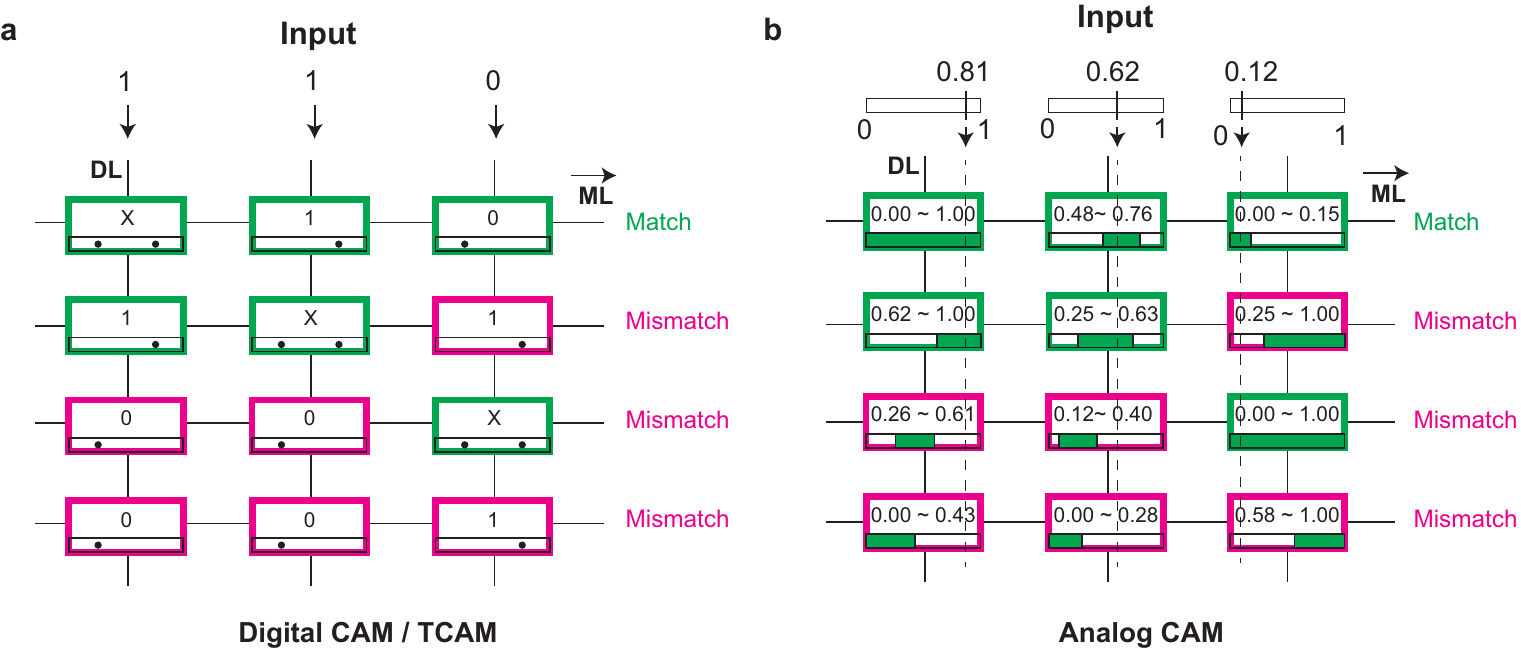}
        \caption{
        \linespread{1.2}\selectfont{}
            \textbf{Schematic of the memristor analog content-addressable-memory concept.}
            \textbf{a,} A digital content-addressable-memory (CAM) compares the input word against all stored words or rows in parallel. Ternary CAM (TCAM) is an extension where in addition to search/stored `0' and `1' values, `X' is a wildcard that always yields a match. Data is searched along vertical Data Lines (DL) and the binary match result of the compare operation between searched and stored words in each row is sensed on horizontal Match Lines (ML).
            The CAM returns the match location of stored data and the searched input (first row here).
            \textbf{b,} The analog CAM searches and stores analog data, where the input data can be a continuous value, and the stored data is a continuous interval with a lower and upper bound representing an acceptance range for a match.
        }
        \label{fig:concept}
    \end{figure}

The proposed analog CAM concept is illustrated in \Mfigure\ref{fig:concept}, where analog voltage values are input to the analog CAM to be searched against the analog ranges encoded by multilevel conductances in the memristors. 
    This is distinct from all previously reported CAMs (SRAM or memristor-based), where only digital signals are searched and stored (\Mfigure\ref{fig:concept}a).
    Similar to a digital CAM, the `match' signal for each row is generated on the matchline (ML) only when all the inputs for every column match the data stored in that row's memory. 
    In contrast to digital CAMs, each analog CAM cell can match a range of analog input voltages (\Mfigure\ref{fig:concept}b), instead of a digital value. 
    The analog CAM can be configured to match a narrow range of discrete values, and therefore one analog CAM cell is a direct functional replacement for multiple digital CAM cells. 
    In addition, similar to storing a `wild card' or `X' in the TCAM, the proposed analog CAM can also store a range of continuous values, which would otherwise be difficult to implement with digital CAMs/TCAMs, but (as described later) is beneficial in internet packet (IP) routing, and more novel applications in decision trees, associative computing\cite{imani2016remam,guo2013acdimm} and probabilistic computing.

\begin{figure}
        \centering
        \includegraphics{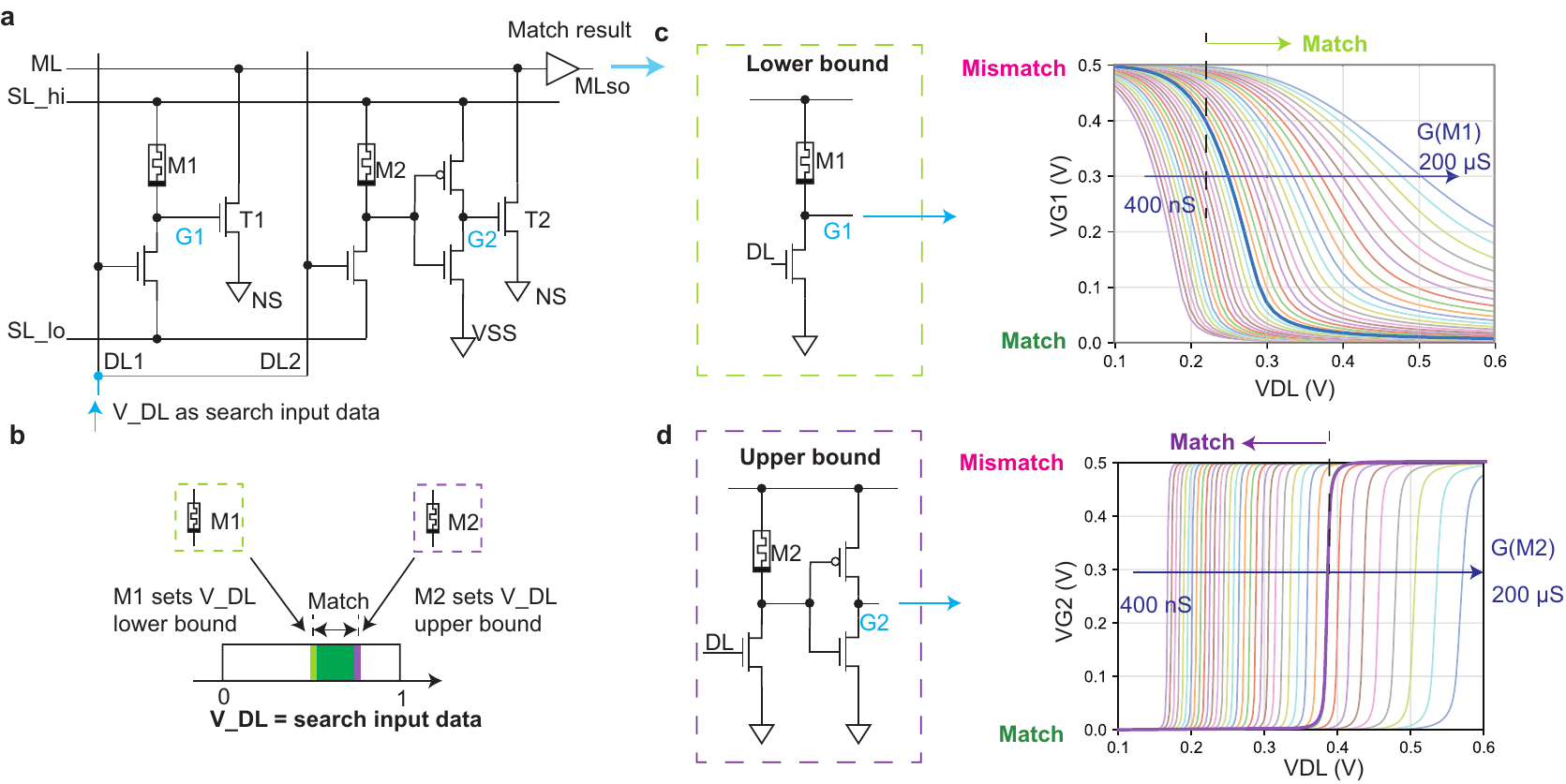}
        \caption{
            \linespread{1.2}\selectfont{}
            \textbf{6-transistors 2-memristors analog content addressable memory circuit. }
            \textbf{a,} Schematic of our proposed analog CAM circuit, composed of six-transistors and two-memristors (6T2M).
            Voltage amplitude on the Data Line (DL) provides search input and the matching result is sensed as the voltage level on the Match Line (ML). 
            \textbf{b,} `Match' result when the analog input is within the range (narrow green band) stored by the cell.
            The stored range is defined by the conductances of two memristors (M1 and M2) in the cell, with M1 determining the lower bound and M2 determining the upper bound of the matching range.
            \textbf{c, d,} Voltage divider sub-circuits translate the input voltage (a search value) to the gate voltage on the ML pull-down transistors. 
            (c) When the input voltage is smaller than the lower bound threshold, the voltage on the gate of the T1 is large enough to pull down the ML, yielding a `mismatch' result. 
            The lower bound threshold is set by the M1 memristor conductance.
            (d) Similarly, when the input voltage is larger than the upper bound threshold, which is tuned by the M2 memristor conductance, the cell returns a `mismatch' result by pulling down the ML. Here, SL\_hi is at  \SI{0.5}{\volt} which sets the max G1 and G2 voltage.
        }
        \label{fig:memristor_cam}
    \end{figure}

    To realize the proposed analog CAM concept, we have designed an analog CAM cell circuit where each cell is composed of six transistors and two memristors (6T2M) (\Mfigure \ref{fig:memristor_cam}a). 
    The analog input search data is mapped to voltage amplitudes V\textsubscript{DL} applied along datalines (DL), and the stored analog range is configured by the programmed conductances of the two memristors of the cell (\Mfigure \ref{fig:memristor_cam}b).
    Similar to existing CAM circuit implementations, the search operation starts by precharging each row's ML to a high logic level, and the MLs stay high (match) only when all of the attached CAM cells of a row match the corresponding input, otherwise discharging and leading to a low logic level (mismatch) on the ML. 
    In the 6T2M design, the ML is connected to pull-down transistors (T1, T2), and is kept high for a `match' result when the gate voltage of the pull-down transistors is smaller than the threshold voltage, keeping the transistor channel in a high resistance state. 

    Each analog CAM cell stores an upper and lower bound for matching against the input search value. These bounds are encoded by two voltage divider sub-circuits which determine the gate voltages of two pull-down transistors connected to the ML. 
    As shown in \Mfigure\ref{fig:memristor_cam}c, the voltage divider sub-circuit consists of a transistor and series connected memristor, which generates the gate voltage (G1) of the pull-down transistor (T1) in the 6T2M analog CAM circuit to embody analog CAM cell's lower bound match threshold.
    When V\textsubscript{DL} is larger than a certain threshold voltage, the transistor is highly conductive and thus the search voltage between SL\_hi and SL\_lo (typically at GND) will mainly drop across the M1 memristor, resulting in a small voltage on G1 that does not turn on the pull-down transistor and yields a match result. 
    The lower bound of the input voltage V\textsubscript{DL} that yields a match is configured by tuning the memristor conductance in the voltage divider. 
    The upper bound of the search range is configured similarly with an independent voltage divider using M2 and an inverter to control the gate voltage (G2) of the second pull-down transistor (T2) (\Mfigure\ref{fig:memristor_cam}b).
    This concept is shown by the simulation of the voltage on G1 and G2 depending on V\textsubscript{DL} with different M1 and M2 memristor conductances (\Mfigure\ref{fig:memristor_cam}c, d).
    As a result, the cell keeps ML high only when V\textsubscript{DL} is within a certain range as defined by the M1 and M2 conductances. 
    As several cells are connected on the same ML in a row, just as in digital CAMs, a row ML outputs `high' only when each cell in the row matches.

    \subsection{Simulations and experiments}
    
    To validate our circuit design and further investigate the memristor-based analog CAM concept, we (1) conducted extensive simulations of individual analog CAM cells and CAM arrays and (2) experimentally measured analog CAM circuit operation in a taped-out silicon test chip. The circuit simulations shown here utilize 16nm design rules to enable projected performance comparisons against current CMOS-based solutions, and our silicon tape-out utilized a 180nm technology node to provide voltage and current overheads and accelerate design to fabrication time.
    
    We first validated the operation of the individual memristor analog CAM cell circuit with circuit simulations (see Methods for details) based on a layout using commercial  \SI{16}{\nano\meter} design rules. The memristor conductance tuning in an analog CAM array is similar to the `write' operation in a 1T1M array and described in the \SInote \ref{not:si_programming}, and \SIfigure\ref{fig:si_memristor_cam_write}, with the parameter summarized in \SItable\ref{tb:si_write}.
    The current design prioritizes feasibility and demonstration of this new circuit concept and is not fully optimized for speed or power consumption.

    \begin{figure}
        \centering
        \includegraphics[scale=0.95]{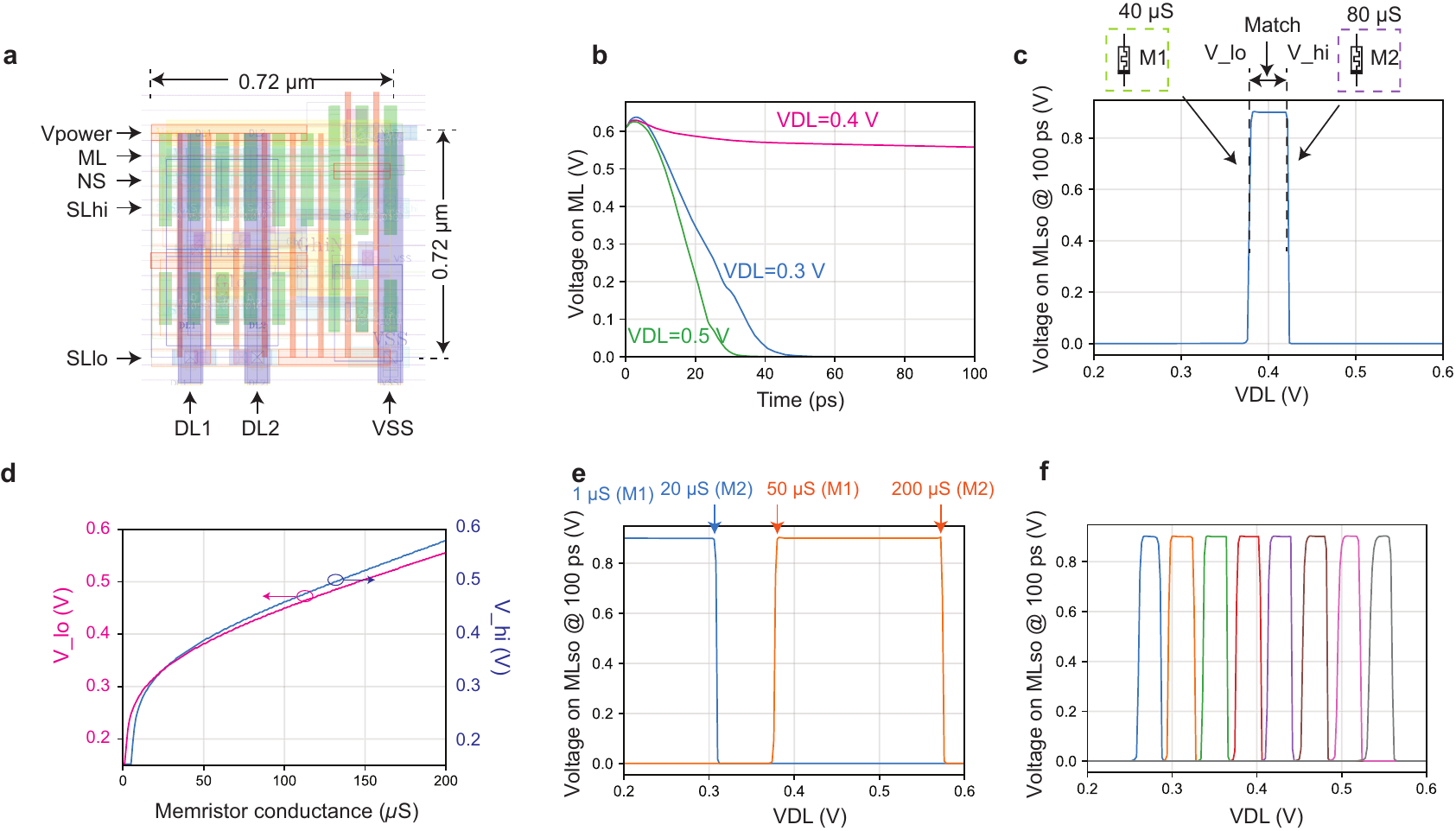} 
        \caption{
        \linespread{1.2}\selectfont{}
            \textbf{Simulations of the memristor analog content-addressable-memory cell.}
            \textbf{a,} The layout design of one analog content-addressable-memory (CAM) cell in an analog CAM array using commercial  \SI{16}{\nano\meter} design rules. 
            The transistors in this proof-of-concept design are over-sized to allow large driving currents.
            \textbf{b,} 
            Simulated transient voltage response on the Match Line (ML) where the different curves show searches with different Data Line (DL) voltages for a matching case (red) and two mismatching cases (blue, green).
            \textbf{c,} The circuit simulation with the same memristor configuration shown in (b) shows that the cell matches a range of DL voltage, whose bounds are independently controlled by the conductances of the two memristors in the cell. 
            \textbf{d,} The simulated relation between the search range and the memristor conductance.  
            The blue curve shows the lower bound of the stored range, while magenta shows the upper bound. 
            \textbf{e, f, } Using differently configured memristor conductances, the cells in the array can store (e) a continuous range of values, or (f) discrete levels (showing eight levels or 3-bits).
        }
        \label{fig:simulation}
    \end{figure}
    
\begin{figure}[!hth]
    \centering
        \includegraphics{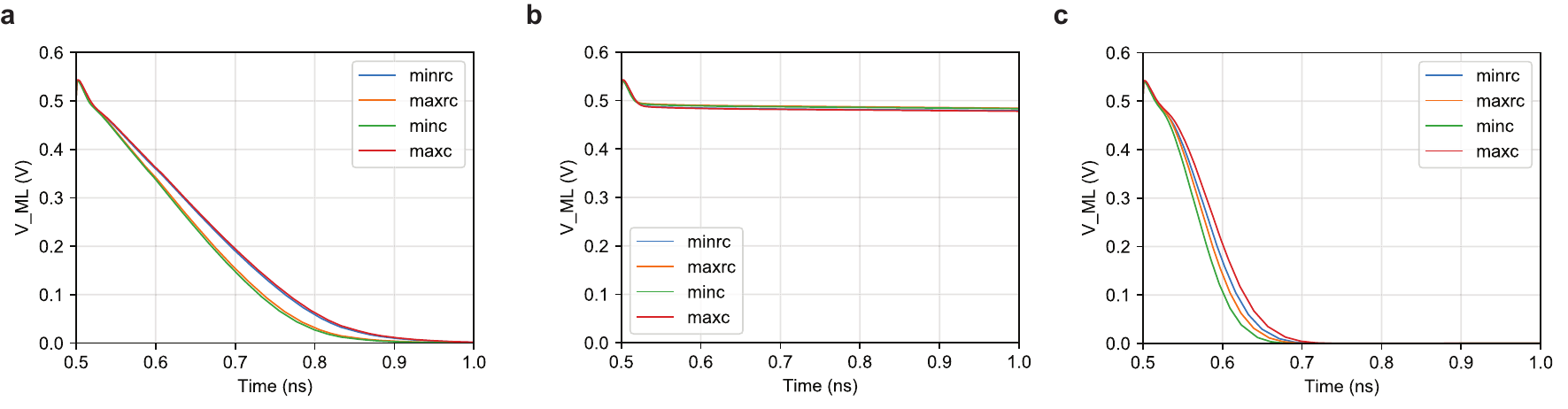} 
        \caption{ \linespread{1.2} \selectfont{}
            \textbf{Simulation of different process corners.}
            Plots show the ML decay during a search in an $86\times12$ analog array with DL voltage of 
            \textbf{a,} \SI{0.3}{\volt} (Mismatch), 
            \textbf{b,} \SI{0.4}{\volt} (Match),
            \textbf{c,} \SI{0.5}{\volt} (Mismatch). 
            The memristor are configured to the same conductance range with that in our single-device simulation in \Mfigure\ref{fig:simulation}b.
        }
    \label{fig:si_simulation_corners}
\end{figure}

    The analog stored value was first configured in the CAM cell by setting the conductance of two memristors in one analog CAM cell to \SI{40}{\micro\siemens} and \SI{80}{\micro\siemens} and apply different $V_\text{DL}$ values to observe the changing $V_\text{ML}$ behavior during the search operation. 
    From simulations we see that after the search is initiated (by pulling SL\_hi high), $V_\text{ML}$ stays high (\Mfigure \ref{fig:simulation}b) when $V_\text{DL}$ is \SI{0.4}{\volt}, indicating a `match', but is discharged low when $V_\text{DL}$ is either \SI{0.3}{\volt} or \SI{0.5}{\volt} for a `mismatch'.
    The operation's timing diagram and voltage parameters are presented in \SIfigure\ref{fig:si_simulation_timing}, from which one sees that the search result can be measured from the transient $V_\text{ML}$ at some time (\eg \SI{100}{\pico\second}) following the search when the voltage difference (i.e. sensing margin) between match and mismatch scenarios is large enough for a sense circuit.
    The simulated ML sensing output at \SI{100}{\pico\second} following the search operation for different $V_\text{DL}$ (\Mfigure\ref{fig:simulation}c) shows that this programmed memristor configuration corresponds to matching for \SI{0.37}{\volt} $< V_\text{DL}<$ \SI{0.42}{\volt}. 
    The lower bound (V\_lo) and the upper bound (V\_hi) of the analog CAM cell's acceptable matching range can be configured independently by tuning the corresponding memristor conductance in the cell. 
    Using the resulting mapping between the voltage bound and conductances (\Mfigure\ref{fig:simulation}d) as a guide, we configured analog CAM cells to match various voltage ranges (\Mfigure\ref{fig:simulation}e) or eight narrower ranges  (\Mfigure\ref{fig:simulation}f) for representing 3-bit discrete voltage levels. 
    CMOS process variation effects are studied with a layout-based simulation under different corner conditions, and results (in \SIfigure \ref{fig:si_simulation_corners}) show that while different conditions slightly change the latency and search boundaries, we can still perform a calibration under those specific conditions to achieve the same bit accuracy as our search boundary is programmed with an iterative program-and-verify approach. 
    Therefore, the proposed analog CAM cell implements the desired functionality and can be used to search for discrete levels (encoding at least 3-bits in a single cell) or for arbitrary analog voltage ranges to encode continuous values.

\begin{figure}
        \centering
        \includegraphics{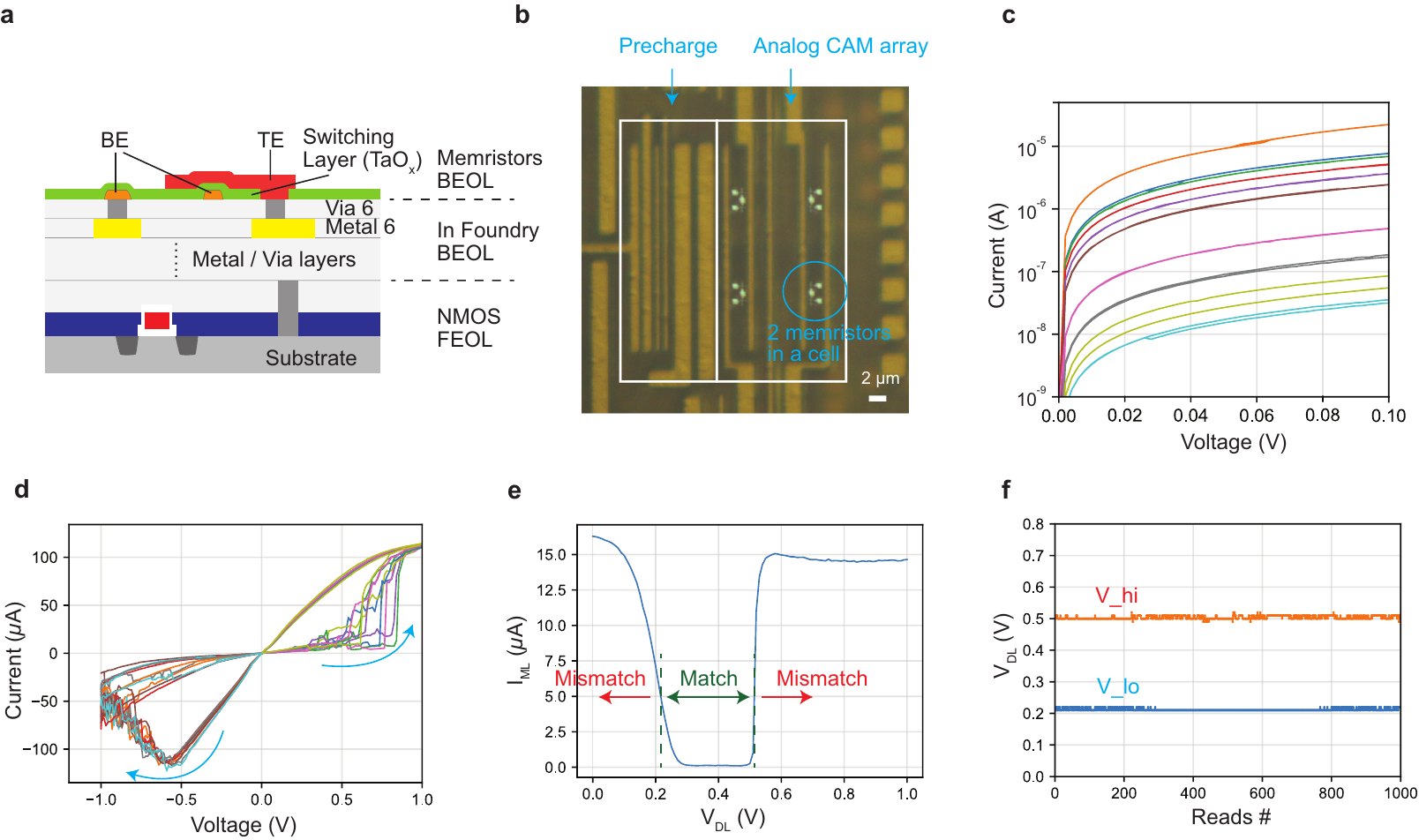} 
        \caption{
        \linespread{1.2}\selectfont{}
            \textbf{Experimental demonstration.}
            \textbf{a,} Cross-sectional diagram showing monolithic integration of memristor on top of CMOS circuits. 
            \textbf{b,} An optical microscope image of an integrated analog CAM array with an on-chip precharging peripheral circuit.
            \textbf{c,} Current-voltage relation of the memristor in the analog CAM circuit for many programmed states, showing the wide continuously tunable conductance range.
            \textbf{d,} The conductance can be programmed from a low conductance state to a high conductance state by a large voltage ($>\SI{0.5}{\volt}$) on SL\_hi, and reversed by a large voltage on SL\_lo.
            \textbf{e,} The Match Line (ML) pull-down current with respect to the applied data line voltage, after programming the memristors. 
            \textbf{f,} The analog CAM cell stored range was extracted from the pull-down current in \SIfigure\ref{fig:si_experiment} (and one trace in (e)) where a high pull-down current ($>$\SI{5}{\micro\ampere}) indicates a mismatch. The range stayed unchanged for 1,000 searches over more than 8,000 seconds.
        }
        \label{fig:experiment}
    \end{figure}

\begin{figure}[!hth] 
    \centering
        \includegraphics{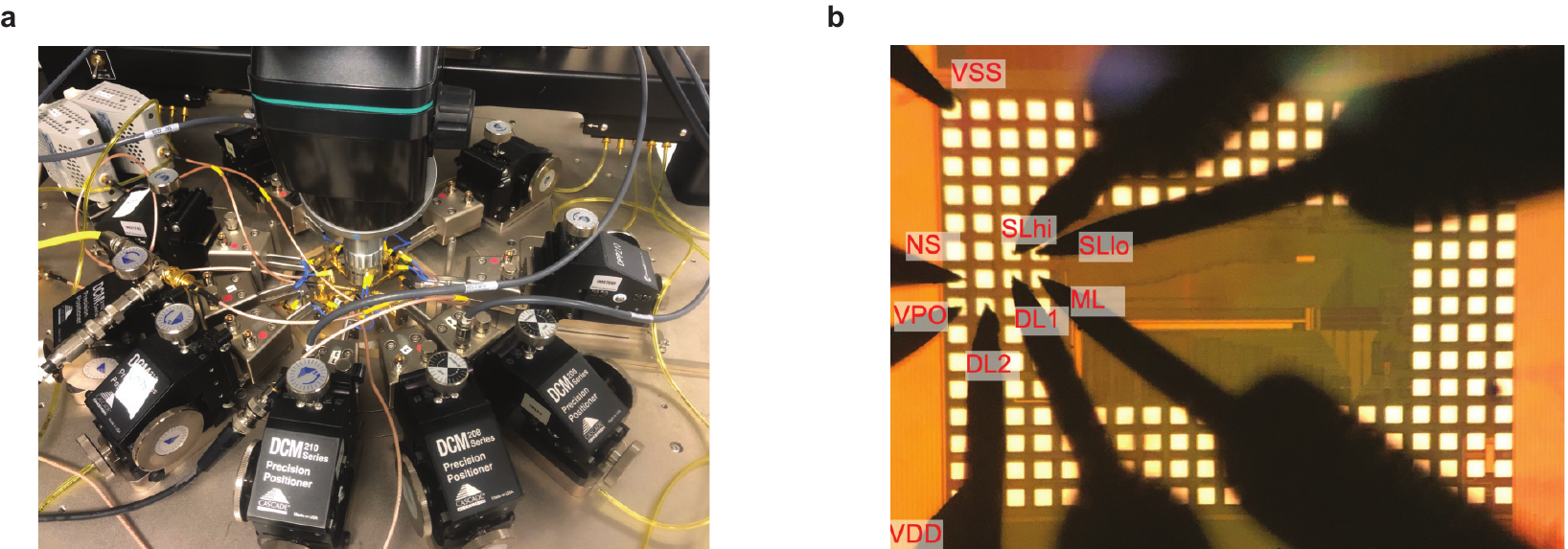}
        \caption{ \linespread{1.2} \selectfont{}
            \textbf{The experimental measurement setup. }
            \textbf{a,} The measurement setup with nine probe manipulators on a probe station. 
            \textbf{b, } Nine probes landed on the chip under measurement. Each probe is connected to either a source measurement unit (SMU) on Keysight B1500 or a direct-current (DC) voltage supply. 
        }
    \label{fig:si_experiment_setup} 
\end{figure}

\begin{figure}[!hth]
    \centering
        \includegraphics{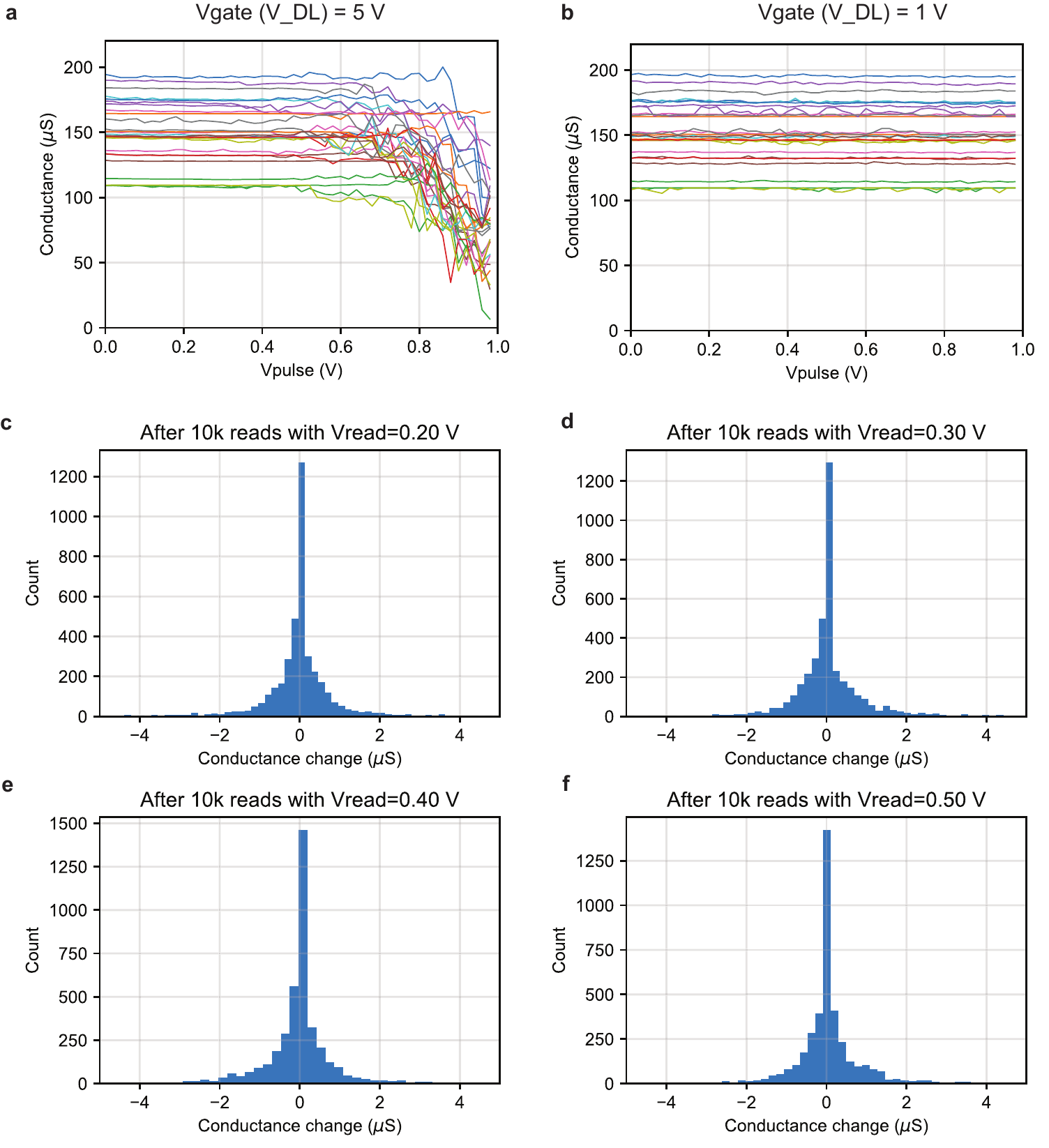}
        \caption{ \linespread{1.2} \selectfont{}
                \textbf{Device stability tests with different voltages on SL\_hi. } 
                \textbf{a, } 
                The memristors are RESET by voltage ($>$0.5 V) pulses applied to SL\_hi with the series transistors fully turned ON. The initial value of the memristors was programmed between 100-200 $\mu$S.
                \textbf{b, } On the other hand, the memristor states were not changed by the applied pulses when in the search mode where the V\_DL is always smaller than \SI{1}{\volt}.
                \textbf{c-f, } The device stability after read operations with different reading voltages. Each panel shows the distribution of the memristor conductance change after 10,000 repeated read operations with the read voltage specified in the title. The conductances do not show noticeable disturb by the read operations, within the noise of the read operation.
        }
    \label{fig:si_devices_vreads}
\end{figure}

\begin{figure}[!hth]
    \centering
        \includegraphics{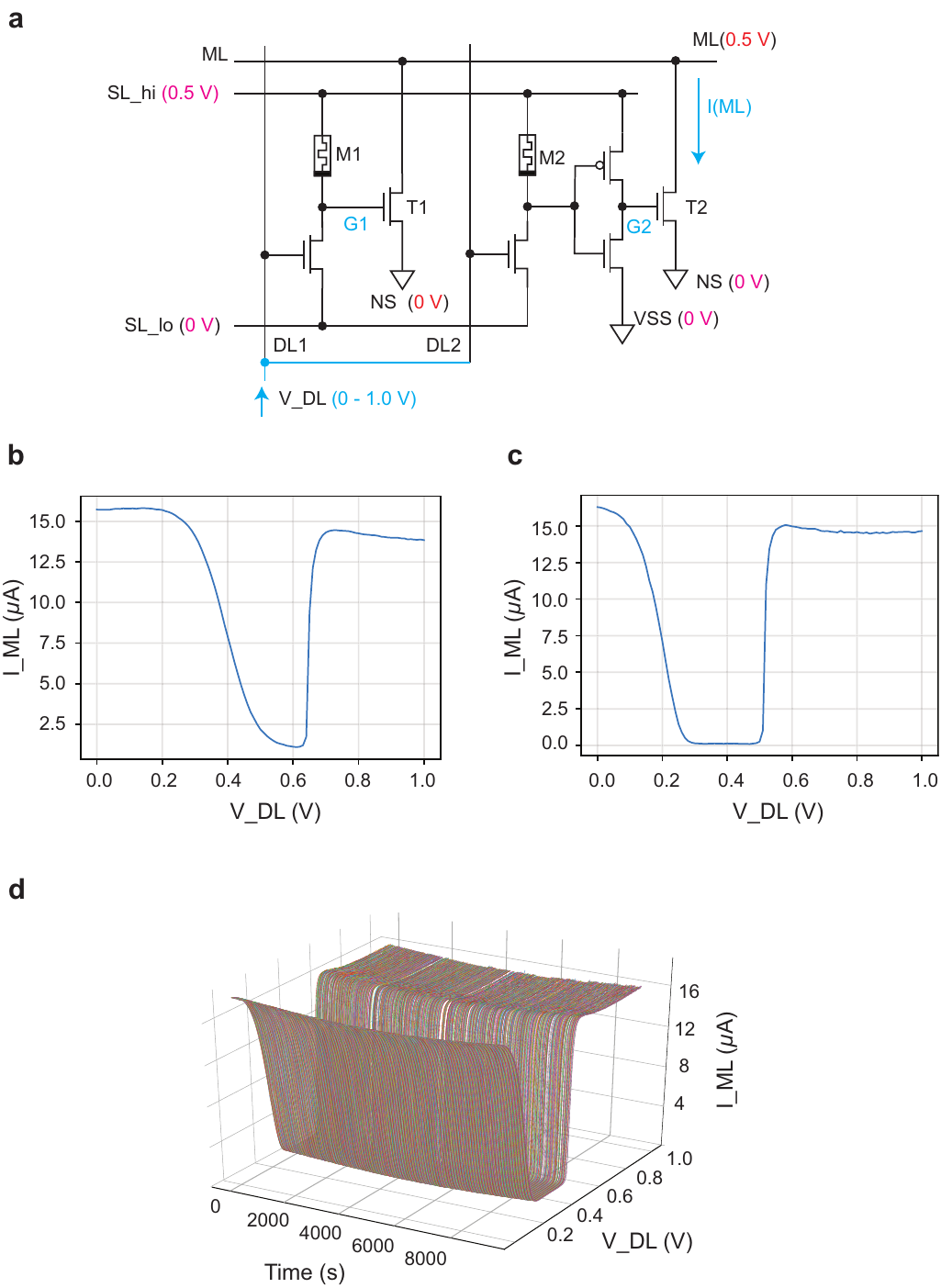}
        \caption{ \linespread{1.2} \selectfont{}
            \textbf{Additional information on the experimental measurement. 
            }
            \textbf{a,} The schematic shows the measurement setup when measuring the relation between the ML discharge current (I(ML)) and the voltage on DL. 
            The magenta labels mark the DC voltage applied to each node. 
            \textbf{b, c, } The ML discharging current with respect to the DL voltage for the two analog CAM cells programmed to search for different ranges. 
            \textbf{d, } The retention / reliability test shows that the cell maintains the searching range for more than 8,000 seconds with 1,000 individual measurements.
            The extracted stored range is shown in \Mfigure\ref{fig:experiment}f.
        }
    \label{fig:si_experiment}
\end{figure}

\begin{figure}[!hth]
    \centering
        \includegraphics{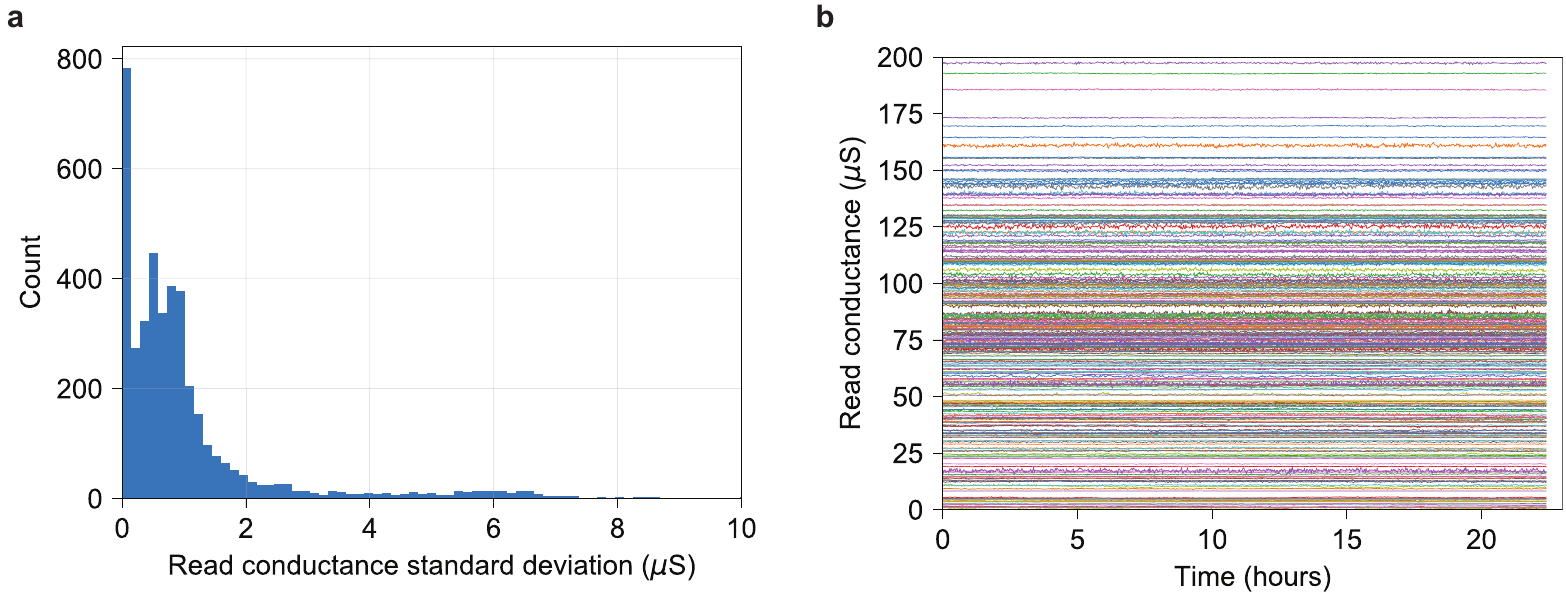}
        \caption{ \linespread{1.2} \selectfont{}
                \textbf{Analog conductance state stability and retention.} 
                \textbf{a, } 
                Distribution of the memristor device read stability shows that the measured (read) conductance for the majority of the memristor devices has a standard deviation of several $\mu$S or smaller. The data was generated from conductance reads with 0.2 V read voltage from all devices in a 64$\times$64 array for 10,000 times. The lateral size of the memristor is 50 nm$\times$50 nm. 
                \textbf{b, } 
                Multilevel retention performance of our integrated \ce{TaO_x} device shows the device conductances did not drift for over 20 hours under room temperature. Each datapoint is averaged from 50 repeated reads.
        }
    \label{fig:si_device}
\end{figure}

    Next, we experimentally verified the proposed analog CAM operation in 6T2M analog CAM cells designed and fabricated at a 180 nm technology node on a silicon test chip (see Methods).
    Memristors of size \SI{50}{\nano\meter}$\times$\SI{50}{\nano\meter} and based on Ta/\ce{TaO_x} were monolithically integrated in a Back End of the Line (BEOL) process with CMOS circuits on top of metal 6 and tungsten vias (\Mfigure\ref{fig:experiment}a). \Mfigure\ref{fig:experiment}b shows the top view image of the analog CAM array with integrated memristors.  
    The integrated memristors have a wide $\sim 10^3$ range of conductance tunability (\Mfigure \ref{fig:experiment}c) and a programming voltage $<$\SI{1}{\volt} under direct-current (DC) sweeps (\Mfigure \ref{fig:experiment}d shows a typical switching curve, and \SIfigure\ref{fig:si_experiment_setup} shows a test chip under measurement), enabling us to validate the search operations experimentally.
    Given that the switching voltage of the memristor device exhibits a certain degree of variation,
    some devices may require a larger voltage to program than others (see \SIfigure\ref{fig:si_devices_vreads}a). Therefore, it may impose challenges for a future technology node which supplies a voltage smaller than \SI{1}{\volt}.
    We programmed two analog cells in the same row to store different ranges using an iterative program-and-verify approach. 
    We observe the match line (ML) pull-down current as we sweep the corresponding data line (DL), with the experimental configuration schematic in \SIfigure\ref{fig:si_experiment}a. 
    As expected, the pull down current is low only when the applied data line voltage falls into the programmed range, indicating a match (\Mfigure\ref{fig:experiment}e), and the other cell attached to the same ML shows a different search range (\SIfigure\ref{fig:si_experiment}b,c).
    \Mfigure\ref{fig:experiment}f shows 1,000 repeated measurements without any observed disturb effects in the stored range (see also \SIfigure\ref{fig:si_experiment}d) and \SIfigure\ref{fig:si_device} and \ref{fig:si_devices_vreads} show the stability statistics of the analog conductance states of the memristors. These results suggest our analog CAM does not require static power to store a range table, nor frequent updates once programmed. 

    The relationship between an analog CAM cell's stored range for a match and the programmed memristor conductances can be understood by the series connected transistor and memristor voltage divider (see \Mfigure\ref{fig:memristor_cam}c, \ref{fig:memristor_cam}d). 
    During a search operation, the serial transistors in the divider are working in the triode regime, as the voltage drop across the transistor channel is fairly small. Under this condition, $V_\text{ML}$ stays high when $V_\text{DL}$ follows \Mequation \ref{eq:vdl_g}, with bounds from the lower bound M1 voltage divider and the higher bound M2 voltage divider.
    \begin{equation}
        \label{eq:vdl_g}
        G_\text{M1} \cdot (V_\text{SLhi}/V_\text{TH,ML} - 1)/\beta + V_\text{TH} \
        \leq V_\text{DL} \
        \leq G_\text{M2} \cdot (V_\text{SLhi}/V_\text{TH,inv} - 1)/\beta + V_\text{TH}
    \end{equation}
    where $V_\text{TH}$, $V_\text{TH,ML}$, $V_\text{TH,inv}$ are the threshold voltages of the transistor in the M1 voltage divider, the T1 pull-down transistor, and the inverter respectively. $\beta$ ($=\partial G_\text{T} / \partial V_\text{DL}$) is a constant coefficient in the transistor transfer function.
    $G_\text{M1}$ and $G_\text{M2}$ are the memristor conductances, which are linearly related to the accepted $V_\text{DL}$ for a match according to the above equation. 
    This analysis is consistent with the results shown in \Mfigure\ref{fig:simulation}d when the transistor is working in the linear regime, but when the memristor conductance is very small (i.e. the transistor voltage drop is small) a numerical simulation is required to extract the precise relation. 
    In practice, we use the simulated relation in \Mfigure\ref{fig:simulation}d to determine the programming target memristor conductance values from the desired stored analog value or range.
    Under this assumption, we scale up our simulations from single cells to large analog CAM arrays to predict performance.
    
    \subsection{Memristor analog content-addressable-memory arrays}
\begin{figure}
        \centering
        \includegraphics[scale=0.95]{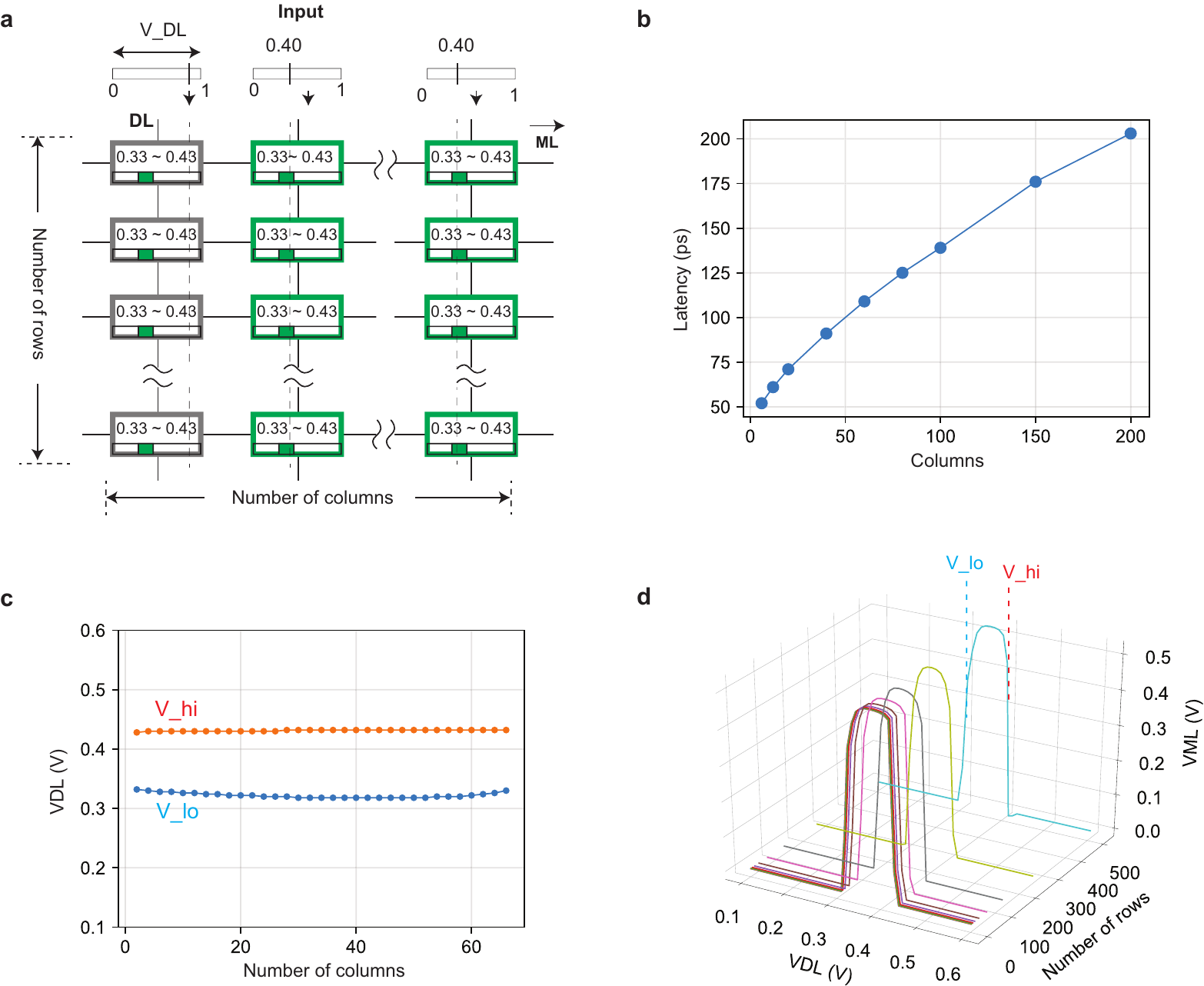}
        \caption{
        \linespread{1.2}\selectfont{}
            \textbf{Analog content-addressable-memory arrays.}
            \textbf{a,} Simulated analog content-addressable-memory (CAM) arrays with different sizes using extracted parasitic parameters. 
            \textbf{b,} The search latency in the one-bit mismatch case increases with the number of columns due to the increased ML capacitance. 
            \textbf{c,} There is a noticeable change in the acceptable search range with an increasing number of columns. 
            In the worst case, up to
            \SI{20}{\milli\volt} change is observed in an array of 64 columns.
            This still allows the capability to store and search 4-bits of information per analog cell across the 64 columns, but going beyond this would be challenging with the present circuit. 
            \textbf{d,} On the other hand, the change with increasing number of rows is negligible up to 512 rows.
            The three-dimensional (3-D) plot shows the sensed ML voltage with respect to the input voltage applied to the DL in arrays with an increasing number of rows (or words). 
            High logic level on the ML indicates that the searching input matches the stored memory, where the lower and the higher bounds ($V$\_lo and $V$\_high) are labeled in the plot. 
            With an increasing number of rows
            the searching range of one cell in the array stays stable. 
        }
        \label{fig:analysis}
    \end{figure}

    While we have demonstrated single analog CAM cell operation in a small array, it is crucial to investigate whether large arrays can be operated without degradation to the desired search operation result. Using extracted parasitic parameters from the 16 nm layout, we constructed analog CAM arrays with arbitrary numbers of rows and columns (see Methods) to study how the analog CAM performs with increasing array size.
    \Mfigure\ref{fig:analysis}a shows the simulation configuration, where the two memristors in each of the analog CAM cells are programmed to \SI{20}{\micro\siemens} and \SI{80}{\micro\siemens} in order to accept  $V_\text{DL}$ from \SI{0.33}{\volt} to \SI{0.43}{\volt}.
    All DLs are biased to \SI{0.4}{\volt}, except for one column DL that is swept from \SI{0.0}{\volt} to \SI{1.0}{\volt} to observe how $V_\text{ML}$ changes. 
    This single-bit mismatch is the worst-case scenario as it represents the situation where the mismatch $V_\text{ML}$ decay is the closest to the match $V_\text{ML}$ behavior.
    Since all cells with $V_\text{DL}=$\SI{0.4}{\volt} match, the $V_\text{ML}$ drop leading to a `mismatch' is initiated by the cells in the column with the sweeping DL.
    Similar to a conventional CAM, ML discharging latency increases with the number of columns because of larger ML capacitances (\Mfigure\ref{fig:analysis}b).
    With increasing number of rows, on the other hand, the latency only changes 5\% in our simulation for an array with 512 rows, suggesting our analog CAM supports the  search for many entries in parallel. 
    
    As expected even with conventional CAMs, increasing the number of columns can also lead to a degradation of the $V_\text{ML}$ (\Mfigure\ref{fig:analysis}c) such that the acceptable search range is slightly changed. 
    Our analysis shows that the degradation of $V_\text{ML}$ with increasing column number from the sub-threshold leakage current of the pull-down transistors (see \SInote\ref{not:si_word_length} for more details), and it can be improved with some emerging devices with extremely low leakage current and sharp transition, as we preliminary explored in \SInote\ref{not:si_emerging}, \SIfigure\ref{fig:si_word_schematic}, \SIfigure\ref{fig:emerging}, and \SIfigure\ref{fig:si_emerging}. 
    Nevertheless, the change in the accepted voltage range is within \SI{0.02}{V} and is thus sufficient to separate more than 20 discrete levels between 0.2 V and 0.6 V for 4 bit searching capability in a 64-column analog CAM array.
    On the other hand, analog CAM arrays with two columns but an increasing number of rows show little change (\Mfigure\ref{fig:analysis}d) in $V_\text{ML}$ with additional rows (simulated up to 512 rows), demonstrating negligible row-wise interference. 
    
    \subsection{Applications}
    The direct advantages of our proposed analog CAM are the improvements in energy and area over existing digital approaches. 
    To demonstrate the potential scale of these improvements, we compared our analog CAM approach with the digital-equivalent for the usecase of classifying Internet protocol (IP) packets, which is a common commercial application for CAMs\cite{pagiamtzis2006tutorial}. 
    The ternary wildcard `X' capability of TCAMs is frequently used to compress multiple table entries into fewer rows in the IP routing look-up table, owing to the fact that most classifying ranges are continuous. 
    With our proposed analog CAM's ability to store broad ranges, this look-up table can be further compressed. 
    Analysis in the previous section suggested that one analog CAM cell is capable of searching 8-64 discrete levels, depending on the size of array and the specific implementation. Columns can therefore be combined as fewer cells are required to store the same amount of information. Additionally, by taking advantage of the range storage capability, fewer rows are required than in a digital CAM/TCAM representation. 
    A real example is given in
    \SInote\ref{not:si_range_cell} and \SIfigure\ref{fig:si_range_search}, which shows a 14x reduction in number of required cells from conventional TCAM to 16-level analog CAM cells, with further reductions possible with improved analog CAM cells. 
    As an area improvement, we estimate an 18.8$\times$ reduction for our analog CAM \SI{12.5}{\square\micro\meter} table compared to an SRAM implementation \SI{235.2}{\square\micro\meter} (see \SIsection\ref{not:si_range_cell} for details).

    To evaluate the search energy improvement with an analog CAM, we simulated the circuit current from all the power supplies with an 86$\times$12 analog CAM array. 
    For a practical evaluation of all digital applications, we custom designed a digital-to-analog (DAC) converter, which imposes additional overhead in both chip area and energy, but our analyses in \SInote\ref{not:si_dac}, \SIfigure\ref{fig:si_dac} and \SItable\ref{tb:si_energy} show this overhead is limited to approximately 10\%.
    The cumulative consumed energy is calculated by integrating the voltage and current over the 16-cycle search with all MLs discharging in the search for the worst-case scenario. 
    Estimating the full array power (including drivers and unoptimized peripherals such as the custom-designed digital-to-analog (DAC) converters), the average total energy per search is $\sim$ \SI{0.52}{\femto\joule} per analog cell, or \SI{0.037}{\femto\joule} for the equivalent number of TCAM bits implementing the same function (see \SInote\ref{not:si_range_cell} for details). 
    The energy per cell consumption is significantly smaller than an SRAM TCAM (0.165 fJ)\cite{huang2010sram65}, which utilizes numerous power saving techniques that provides a $>10\times$ reduction but are not implemented in our analog CAM yet, and a conventional memristor TCAM (0.17 fJ)\cite{graves2019TNANO}. 

    In addition to serving as a higher data-density digital replacement, the proposed analog CAM offers novel applications when the range search capability is utilized.
    As an example, a decision tree with binary and non-binary classification features can be implemented in the analog CAM directly by mapping each root to leaf path to a row in the analog CAM (see \SIfigure\ref{fig:si_dt} and \SInote\ref{not:decision_tree} for mapping details). 
    Logically, each root-to-leaf path traverses a series of nodes with Boolean ANDs between elements in a given input feature vector (\SIfigure\ref{fig:si_dt}a). Since AND is commutative, we can reorder the nodes such that feature variables are processed in the same order for all paths\cite{buschjager2017decision}. Nodes for the same feature are combined into one node and ``don't care'' nodes can be inserted for features absent from a specific path, such that each path is of equal length. This representation can then be directly mapped to the analog CAM array, with each root to leaf path a row. As the matching row can directly drive the readout of the  classification result, tree traversal becomes a one-cycle operation (\SIfigure\ref{fig:si_dt}b). This high throughput and low latency operation are highly advantageous and differentiated from current usage. While ensemble tree-based models are a popular state-of-the-art machine learning approach for classification and regression across diverse real-world applications, these models are difficult to optimize for fast runtime without accuracy loss in standard architectures\cite{tracy2016rf} due to non-uniform memory access patterns, resulting in unpredictable traversal and classification times today. With our proposed analog CAM, it becomes feasible to process large tree-based models at high data rates, such as those required for streaming applications or autonomous vehicles.

    \subsection{Discussion}

    In summary, we have proposed an analog CAM cell circuit taking advantage of the analog memristor conductance tunability for the first time. 
    A practical circuit implementation composed of six transistors and two memristors has been demonstrated in both experiment and simulation. 
    The analog CAM increases memory density significantly, as one analog CAM cell can store multiple bits with only six transistors while an SRAM CAM cell stores 1 bit values with 10 transistors, or ternary values with 16 transistors in a TCAM cell. 
    The analog capability opens up the possibility for directly processing analog signals acquired from sensors, and is particularly attractive for Internet of Things applications due to the potential low power and footprint. The output of the analog CAM after the sense amplifiers is digital, and thus can also remove the analog-digital conversion cost entirely. 
    Finally, the functionality of our analog CAM with interval storage is intrinsically different from digital CAMs, which may enable new computing applications in decision tree models, associative computing and probabilistic processing where inexact compares and real-valued analog transition probabilities are common.

\begin{figure}[!hth]
    \centering
        \includegraphics[scale=0.93]{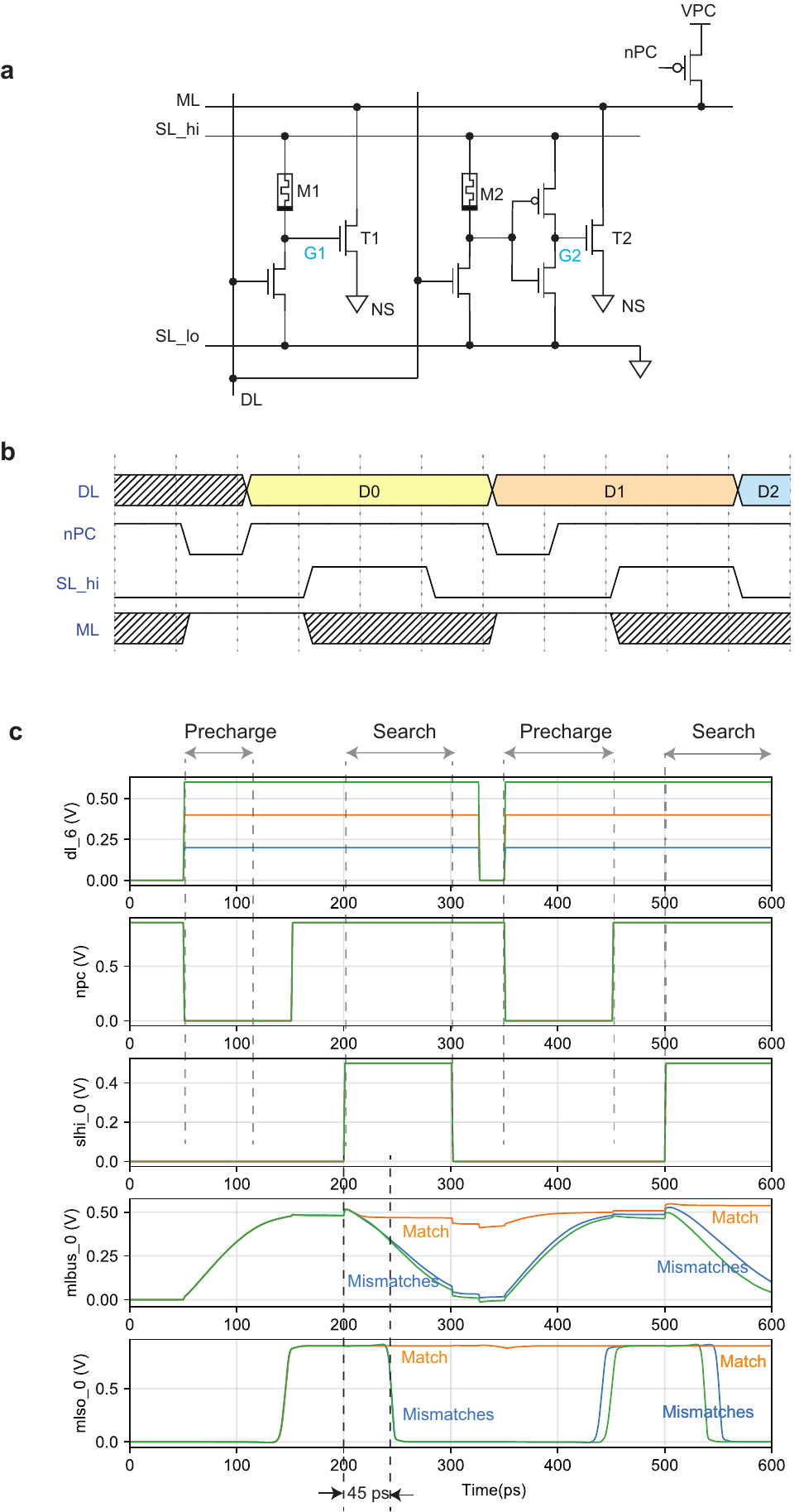} 
        \caption{ \linespread{1.2} \selectfont{}
            \textbf{Transient circuit simulation for the timing of the search operation.}
            \textbf{a,} A simplified schematic of one analog CAM cell with precharging p-type MOSFET attached to its ML. 
            \textbf{b,} The timing diagram for a search operation, where ML precharging is initiated by setting PC high, and the search operation by SL\_hi.
            \textbf{c,} Simulated transient plot of the precharging and search operation in a 86$\times$12 analog CAM array for two cycles. The ML is pulled down within \SI{100}{\pico\second} when the DL voltage mismatches the stored range, and is kept high in the case of a match. 
        }
    \label{fig:si_simulation_timing}
\end{figure}

    \begin{methods}
    
    \subsection{Memristor integration} 
    The memristors are monolithically integrated on CMOS fabricated in a commercial foundry in a \SI{180}{\nano\meter} technology node. 
    The integration starts with a removal of native oxide on the surface metal with reactive ion eaching (RIE) and a buffered oxide etch (BOE) dip. 
    Chromium and platinum are then sputtered and patterned with E-beam lithography as the bottom electrode, followed by reactive sputtered \SI{2}{\nano\meter} tantalum oxide as switching layer and sputtered tantalum metal as the top electrode. 
    The device stack is finalized by sputtered platinum for passivation and improved electrical conduction.

    \subsection{Circuit simulation for analog CAM cell and arrays}
    The proposed 6T2M analog CAM cells designed in the Cadence Virtuoso Custom IC design environment (version ICADV12.1-64b.500.14), and the simulation result is analyzed and post-processed with HP-SPICE (version 4.11). 
    The simulations utilize the  TSMC \SI{16}{\nano\meter} library and the designs follow the corresponding rules. 
    The voltage parameters and timing diagram are shown in \SIfigure\ref{fig:si_simulation_timing}. 
    A custom python script generates the netlist for analog CAM arrays with different numbers of rows and columns and arbitrary configured memristor conductances and input voltages. 
    In the netlist, the parasitic parameters are extracted from the taped out layout, including the wire resistance \SI{1.91}{\ohm}, \SI{2.27}{\ohm}, \SI{0.85}{\ohm} per block  for ML, DL, SL, and capacitance \SI{0.227}{\femto\farad}, \SI{0.324}{\femto\farad}, \SI{0.454}{\femto\farad} between different analog CAM cells. 
    The voltage stimulus is always applied to the nodes that are the furthest from the ML sensing node, so that the impact of the wire resistance is the most significant, \textit{i.e.} the worst case scenario. 
    
    \subsection{Electrical characterization}
    The electrical characterization is conducted with a semiconductor parameter analyzer (Keysight B1500A) and Cascade probe station under room temperature. 
    The conductance programming is performed with quasi-static direct-current (DC) sweeps, and where the current through the device is limited and controlled by the series connected transistor. 
    After programming, the memristor conductance is readout by applying a small voltage across the memristor with the series-connected transistor fully turned on. The search operation is conducted by applying and sensing voltages from the corresponding node to/from a source measurement unit (SMU) on the B1500A.
    A custom python code with PyVISA library was used to control the equipment.

    \end{methods}
    
    \section*{Data availability}
    The data supporting plots within this paper and other findings of this study are available with reasonable requests made to the corresponding author.
    \clearpage
    \section*{References}
    \bibliographystyle{naturemag}
    \bibliography{aCAM}

    \begin{addendum}
     \item[Acknowledgement] 
    The authors acknowledge fruitful discussions with Jim Ignowski and Yuanming Zhu.
    This research was based upon work supported by the Office of the Director of National Intelligence (ODNI), Intelligence Advanced Research Projects Activity (IARPA), via contract number 2017-17013000002, and by the Army Research Office under Grant Number W911NF-19-1-0494. 
    The views and conclusions contained in this document are those of the authors and should not be interpreted as representing the official policies, either expressed or implied, of the ODNI, IARPA, Army Research Office or the U.S. Government. The U.S. Government is authorized to reproduce and distribute reprints for Government purposes notwithstanding any copyright notation herein.
     \item[Author contribution] 
    C.L., C.G., J.P.S. contributed to the conception of the idea  and wrote the manuscript. C.L. designed the experiment and collected the simulation and experimental data. X.S. integrated the memristors, D.M. did the layout design, M.F. and C.L. designed the custom digital-to-analog converter. C.G., C.L., G.P., J.P.S. contributed to the idea of mapping the decision tree.
     \item[Competing Interests] The authors declare that they have no competing interests.
     \item[Correspondence] Correspondence and requests for materials
    should be addressed to can.li@hpe.com, catherine.graves@hpe.com, or john-paul.strachan@hpe.com
    \end{addendum}

\beginsupplement

\section{Supplementary Figures}

\begin{figure}[!hth]
    \centering
        \includegraphics{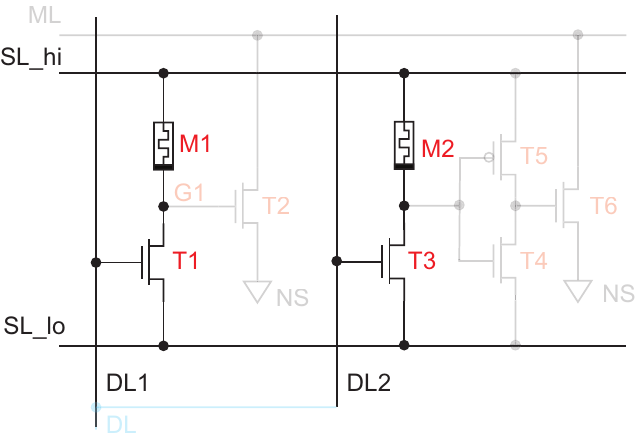} 
        \caption{ \linespread{1.2} \selectfont{}
            \textbf{The schematic for the programming operation of the memristors in an analog CAM cell}
        }
    \label{fig:si_memristor_cam_write}
\end{figure}

\begin{figure}[h!]
    \centering
        \includegraphics{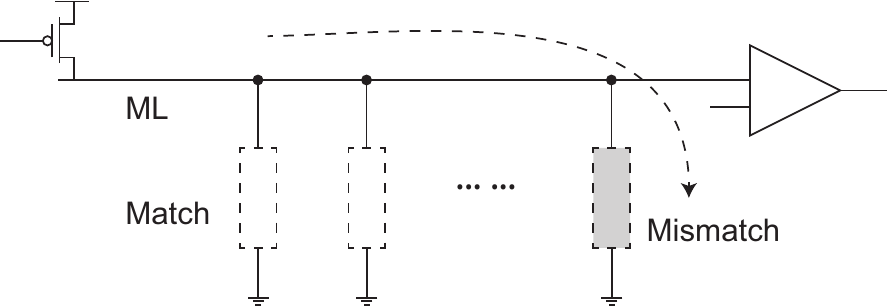} 
        \caption{ \linespread{1.2} \selectfont{}
            \textbf{The schematic of an analog TCAM word in with only one-bit is mismatch during the search operation.}
        }
    \label{fig:si_word_schematic}
\end{figure}

\begin{figure}[!hth]
    \centering
    \includegraphics[scale=1]{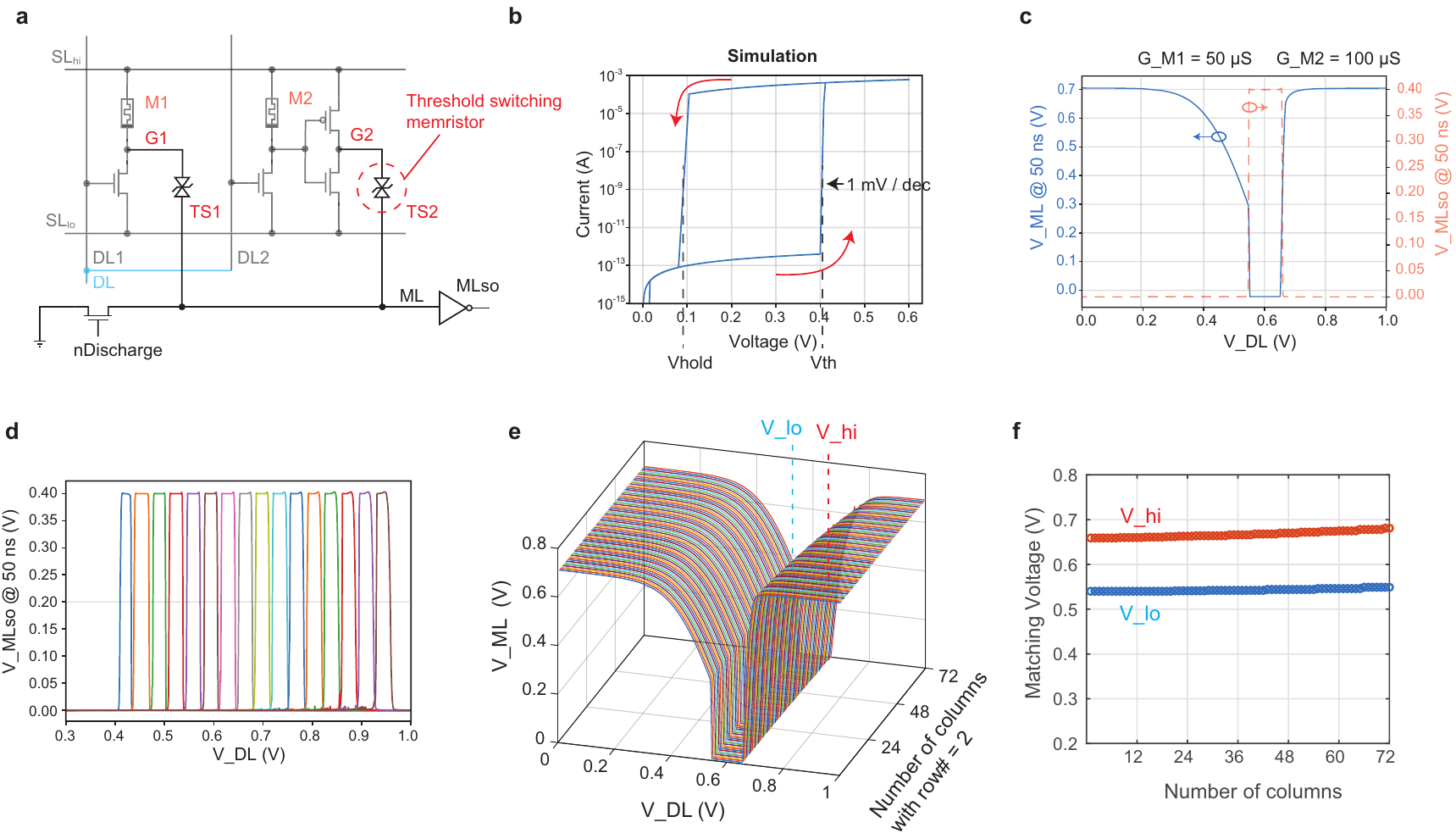} 
    \linebreak
    \caption{
    \linespread{1.2}\selectfont{}
        \textbf{An analog CAM circuit using both volatile and non-volatile memristors}
        \textbf{a, } Schematic for our analog CAM circuit, composed of four transistors and four memristors (two non-volatile and two volatile).
        The volatile threshold switching memristors pull up the ML in the case of a mismatch, which replaces the pull-down transistors in the 6T2M circuit. 
        \textbf{b, } The volatile memristor is a threshold switching device with a very sharp transition between states (\eg \SI{1}{\milli\volt/dec}\cite{s_midya2017am}), therefore reducing column interference issues exposed in earlier simulations.
        \textbf{c, }
        The match line stays low only when the input pattern (V\textsubscript{DL}) matches the stored range. 
        The dashed line shows the signal after the match line sense amplifier output (V\textsubscript{MLSO}), which inverts and converts the analog signal to a binary `match' (high) or `mismatch' (low) signal.
        \textbf{d, } Due to the much smaller $\partial G_\text{pu}/ \partial V_\text{G}$, the cell promises the capability to store more accurate ranges and accordingly more bits of discrete levels (showing 16 levels). 
        \textbf{e, f} The search operation with a simulated array of different word width. 
        The programmed memristor configuration and the expected searching range is the same as that in \Mfigure\ref{fig:analysis}, but the searching range is altered less than $\SI{10}{\milli\volt}$ indicating the capability to store 5-bits of information, which is close to the precision limit of most non-volatile memristor devices.
    }
    \label{fig:emerging}
\end{figure}

\begin{figure}[!hth]
    \centering
        \includegraphics{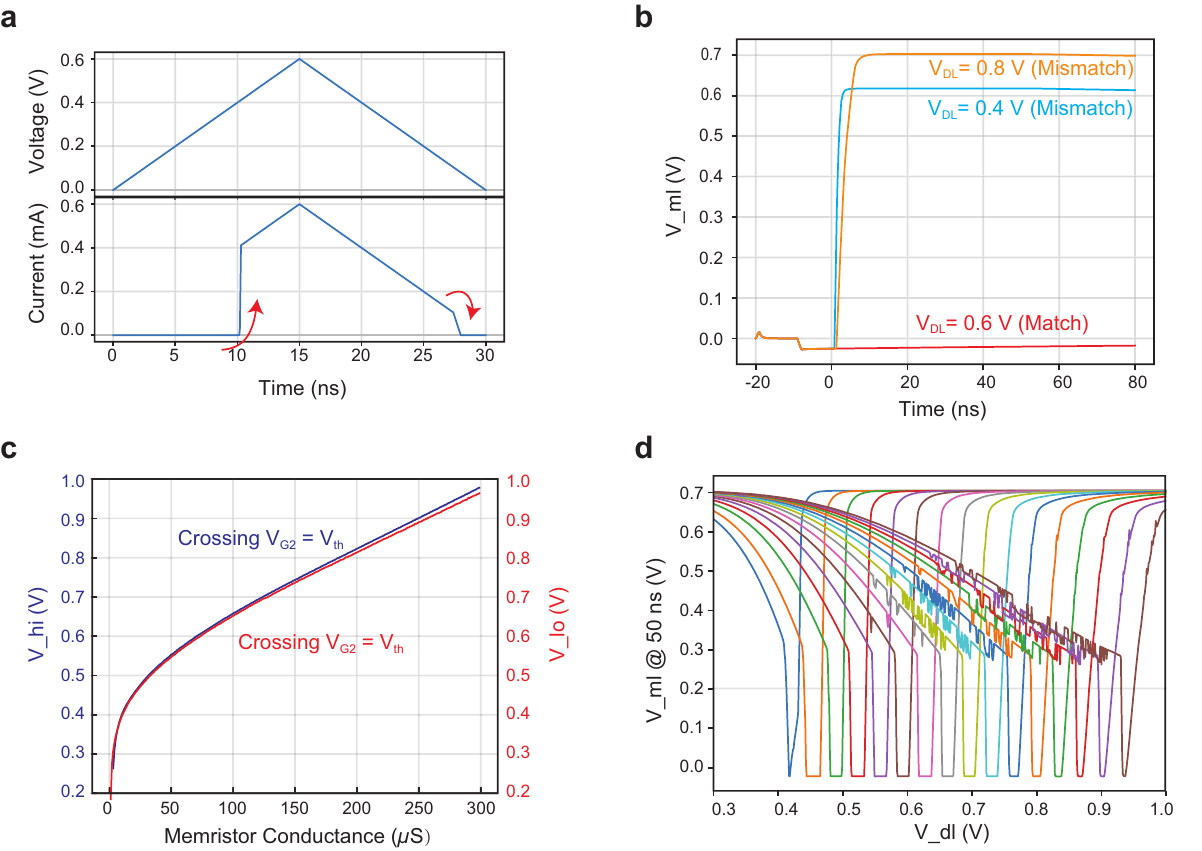} 
        \caption{ \linespread{1.2} \selectfont{}
            \textbf{Additional information on the analog CAM with threshold switching memristor.} 
            \textbf{a, } The transient response of the simulated threshold switching device. The applied voltage is swept from \SI{0}{\volt} to \SI{0.6}{\volt}. 
            The current flowing through the device abruptly increases when the voltage reached the threshold voltage (\SI{0.4}{\volt} in this case), and decreases abruptly after the voltage drops below the hold voltage (\SI{0.1}{\volt} in this case). 
            \textbf{b, } The transient voltage response on the match line (ML) during the search operation, for the cases that the stored range matches (in red) and mismatches (in blue and yellow) the input ($V_\text{DL}$) respectively. 
            \textbf{c, } The relation between the searching ranges and the corresponding memristor conductance.
            \textbf{d, } The match line readout voltage at the time of \SI{50}{\nano\second} after the search operation for differently configured searching ranges.
            The corresponding output after a sense amplifier is plotted in \Mfigure\ref{fig:emerging}d.  
            The plots show the device can be used to store and search at least 16 discrete levles. 
        }
    \label{fig:si_emerging}
\end{figure}

\begin{figure}[!hth]
    \centering
        \includegraphics{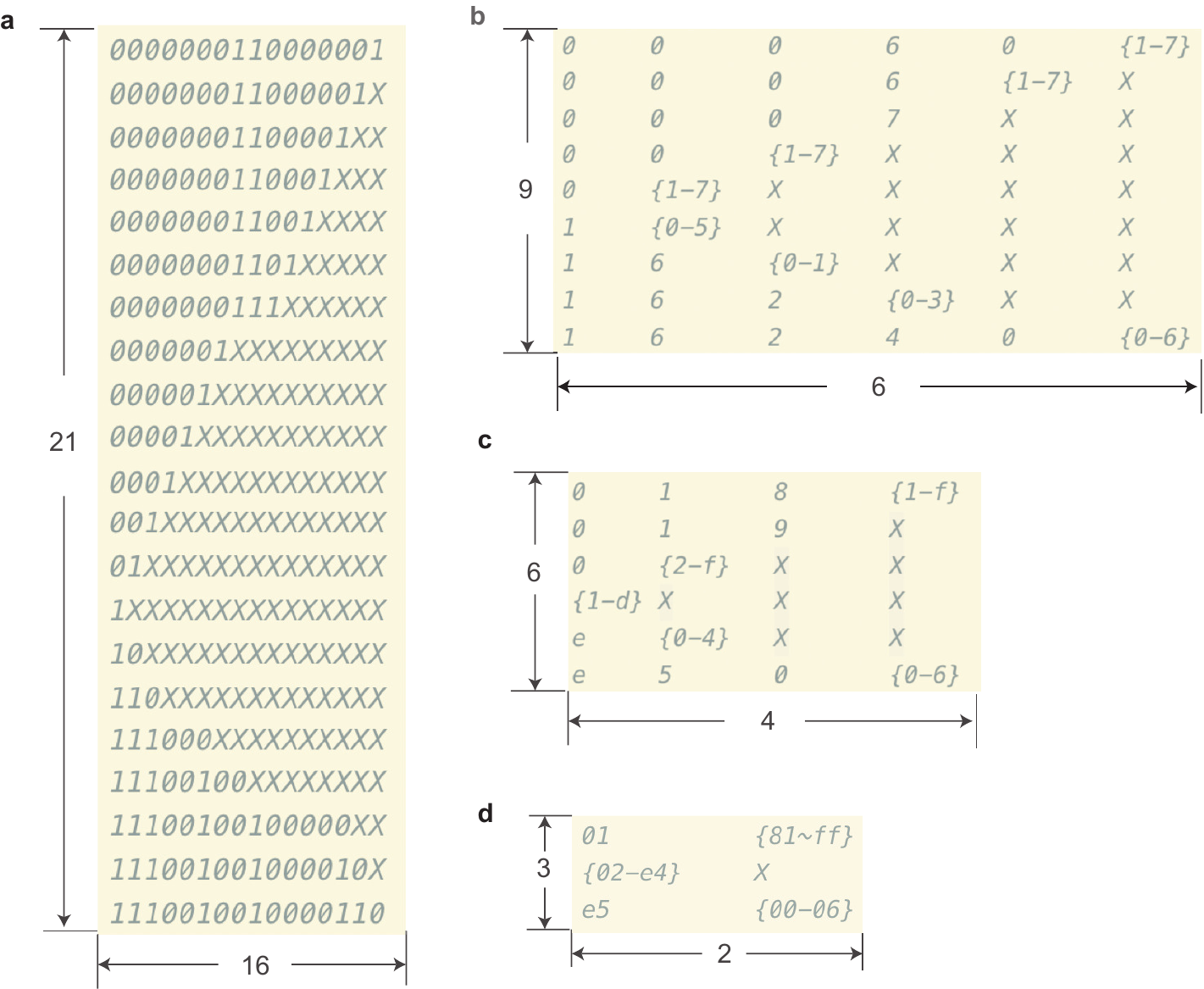} 
        \caption{ \linespread{1.2} \selectfont{}
            \textbf{CAM tables for searching a range between 385 and 58630, with (a) TCAM, (b) 3-bit analog CAM, (c) 4-bit analog CAM and (d) 8-bit analog CAM.}
        }
    \label{fig:si_range_search}
\end{figure}

\begin{figure}[!hth]
    \centering
        \includegraphics{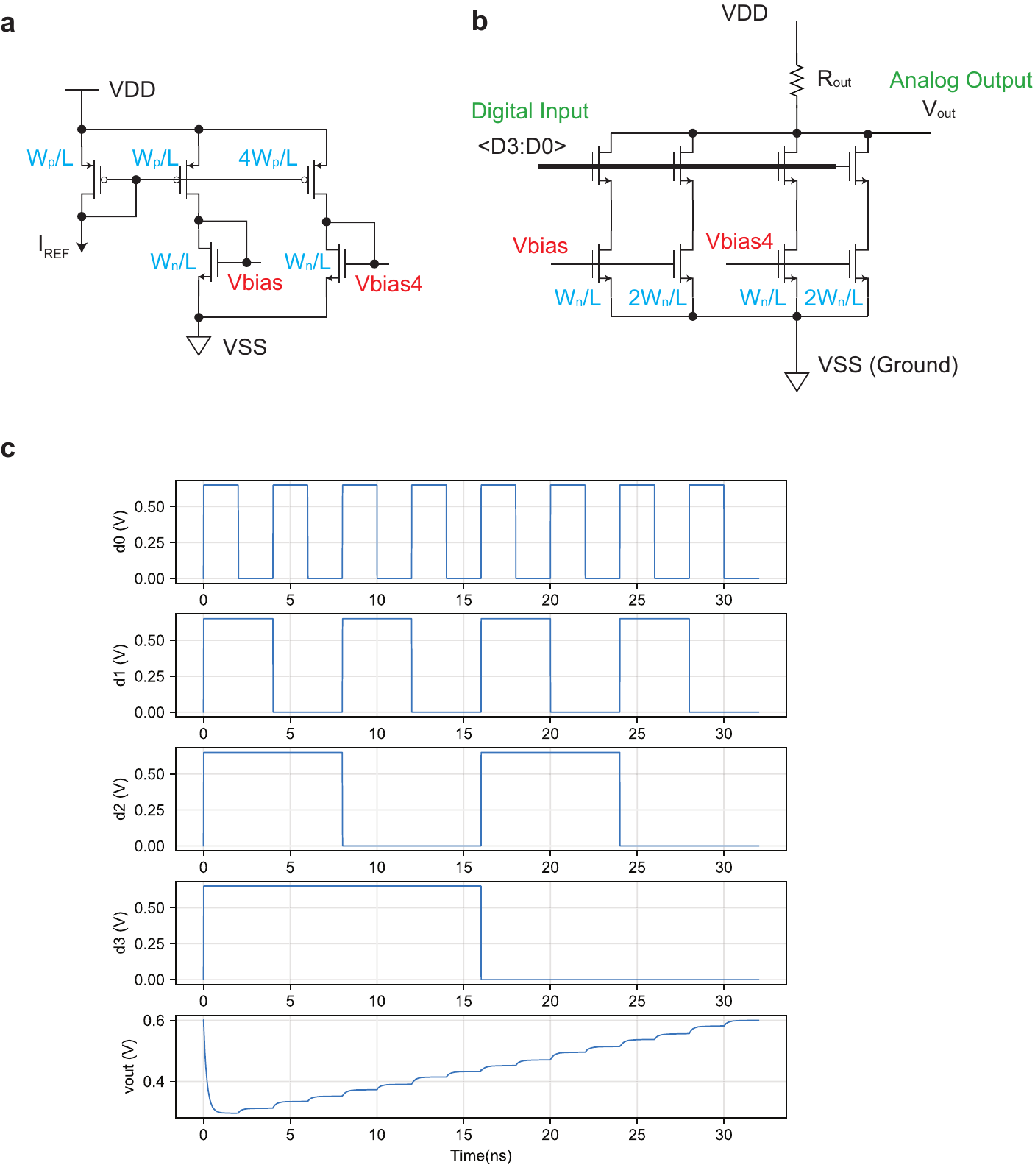} 
        \caption{ \linespread{1.2} \selectfont{}
            \textbf{A DAC design example for analog CAM.}
            The circuit schematic for the modified current-sterring DAC design with \textbf{a,} shared input current mirrors for $I_\text{REF}$ and $I_\text{REF}\times4$ and \textbf{b,} the output current mirrors that converts the digital inputs to the analog output signal. 
            \textbf{c,}
            The simulated DAC operation with different digital inputs.
        }
    \label{fig:si_dac}
\end{figure}

\begin{figure}[!hth]
    \centering
        \includegraphics{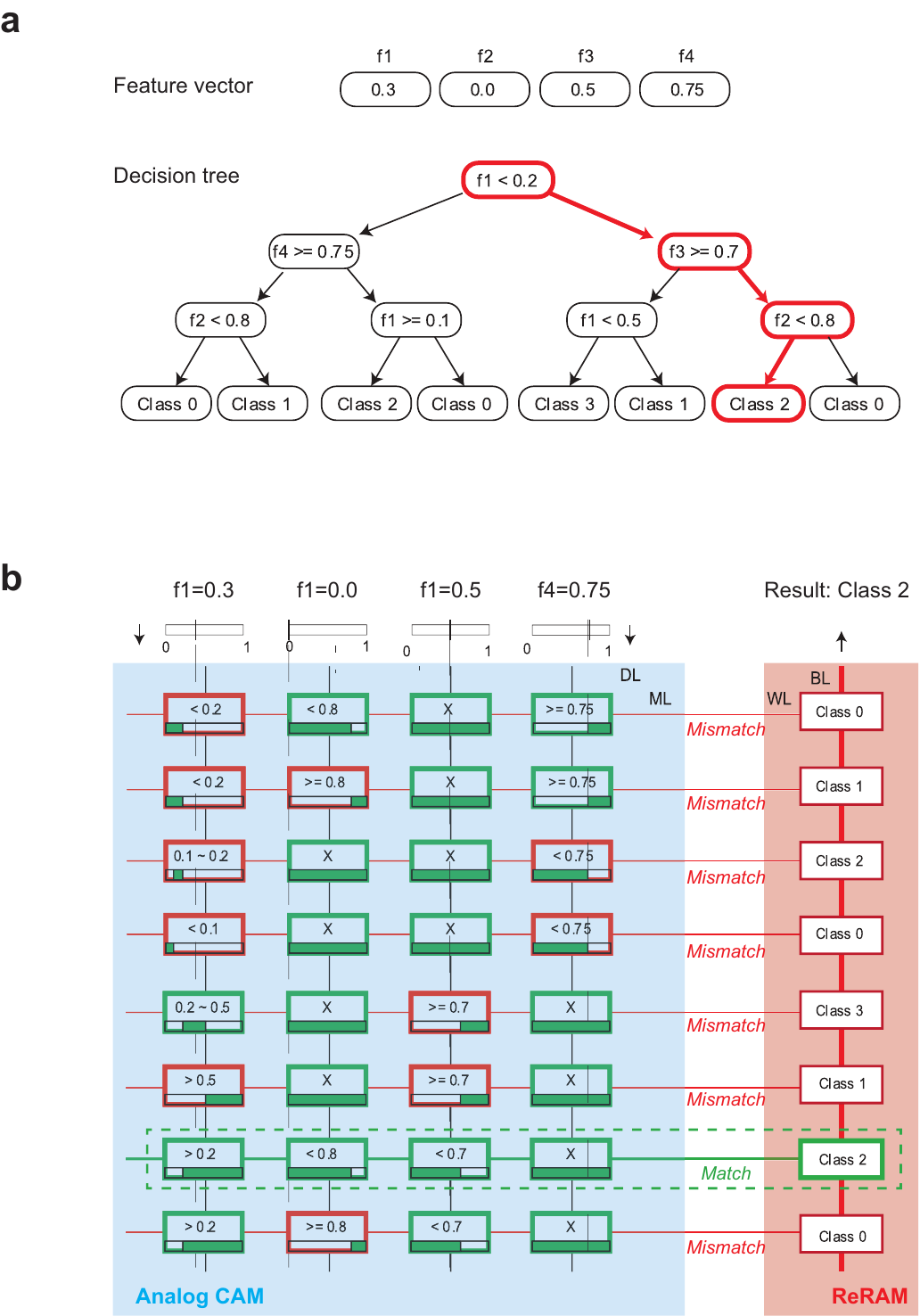} 
        \caption{ \linespread{1.2} \selectfont{}
            \textbf{Decision Tree in an analog CAM and a memristor RAM for fast tree traversal.}
            (a) The tree structure of a sample decision tree directly mapped to a (b) memristor analog content addressable memory (CAM) in conjunction with a memristor random access memory (RAM). 
        }
    \label{fig:si_dt}
\end{figure}

\section{Supplementary Tables}

    \begin{table}[H]
        \centering

        \begin{tabular}{l|llll}
        \hline
        Operation & SL\_hi & SL\_lo   & DL1    & DL2    \\
        \hline
        Set M1    & Vset & 0      & Vg,set & 0      \\
        Reset M1  & 0    & Vreset & V\textsubscript{DD}     & 0     \\
        Set M2    & Vset & 0      & 0      & Vg,set \\
        Reset M2  & 0    & Vreset & 0      & V\textsubscript{DD}    \\
        Read M1   & Vread & 0     & V\textsubscript{DD}     & 0     \\
        Read M2   & Vread & 0     & 0      &V\textsubscript{DD}     \\
        \hline
        \end{tabular}
        \caption{ Write operation of the analog CAM cell}
        \label{tb:si_write}
    \end{table}

    \begin{table}[H]
        \centering
        \begin{tabular}{r|cc}
        \hline
        Supply & Energy/search & Energy/search/cell    \\
        \hline
        ML precharging  & 102.9 fJ    & 0.10 fJ \\
        SLhi driver     & 298.5 fJ    & 0.29 fJ \\
        Others          & 86.4 fJ     & 0.08 fJ \\
        DAC             & 52.1 fJ    & 0.05 fJ \\
        \hline
        Total           & 539.9 fJ    & 0.52 fJ \\
        \hline
        \end{tabular}
        \caption{  Energy break down during searches in a $86\times12$ array }
        \label{tb:si_energy}
    \end{table}

\pagebreak
\section{Supplementary Notes}

\begin{sinote}
    Programming of memristor devices in an analog CAM array
    \label{not:si_programming}

    The memristors in the analog CAM need to be properly programmed before the search operation, and the circuit search operation has been introduced in the main text. 
    \SIfigure \ref{fig:si_memristor_cam_write} shows the schematic operation during memristor programming. The key elements that are involved in the write operation is highlighted in the figure, while other lines not used for writing are greyed out. From the schematic, one sees that the write operation is similar to that of a one-transistor one-memristor (1T1M) array: The DLs (data lines) select the memristor device to be programmed (as well as the aCAM array column), and the programming voltage is applied through the SL\_hi and SL\_lo row wires to set (program the device from a low conductance state to a high conductance state) or reset the the device (and select the aCAM array row). An analog voltage can be applied to DLs to set a compliance current during the set operation for a better multilevel tunability\citesicite{s_li2018analog, s_hu2018dpe}. 
    
    In the case that the programmed memristor conductance needs to be verified after the write operation, the conductance of a given memristor be read out by measuring the current while applying a reading voltage across SL\_hi and SL\_lo, with DL activated to select the device. \SItable \ref{tb:si_write} summarizes the detailed voltage signals required for various operations. 

    Iterative programming of memristors may take a long time to complete and consume significant amount energy. 
    However, the applications proposed in this work do not require frequent updates of memristor conductances and therefore the overhead of programming memristors is negligible.
    In addition, due to the nonvolatility of memristor devices, the analog CAM does not require frequent re-programming or stand-by power once programmed. 
    In the case of applications that do require frequent updates (\eg \textit{ in situ} training of a decision tree), the peripheral circuit design would need to be carefully designed and optimized.
    
\end{sinote}

\begin{sinote}
    Analysis on the effect of word length
    \label{not:si_word_length}
    
    The search operation of our analog CAM can be illustrated in \SIfigure\ref{fig:si_word_schematic}. 
    In the schematic, each dashed square represents a ML pull-down path in one analog CAM cell, which is a transistor whose channel conductance is low when the stored content matches the input and high when mismatches. 
    Therefore, the ML stays high only when all the cells in a row match the given input, and the result is sensed by a sense amplifier attached to the ML.
    However, since all the pull-down paths are connected in parallel, when the word is long enough the overall pull-down conductance for a `match' case could be higher than a mismatch worst case (i.e. a single-bit mismatch), leading to a sensing error and/or changes in the accepted search range of an analog CAM cell.

    The problem is examined by quantitative analysis. The ML discharge process can be modelled by a RC decay or the discharging of a capacitor with a constant current sink (see \SIfigure\ref{fig:si_word_schematic}).
    When all the analog CAM cells match the input, the equivalent conductance of the pull-down path is roughly $N\cdot G_\text{T, OFF}$, where $N$ is the word length (i.e. width of the analog CAM array row).
    The pull-down path conductance for a mismatched case is larger or equal to $G_\text{T, ON} + (N-1)\cdot G_\text{T, OFF}$. 
    A margin is required to differentiate the `match' and the `mismatch' cases, which requires that the match case overall pull-down conductance is larger than that for a mismatch case. 
    If we define the ratio of the conductance difference as $\beta$ ($>1$), the requirement can be described in \SIequation \ref{eq:si_word_length}, and therefore the word length is limited by the conductance ON/OFF ratio (dynamic range) of the pull-down transistor. 

    \begin{equation}
        G_\text{T, ON} > \left[ \left(\beta-1\right)N+1 \right] \cdot G_\text{T, OFF}
        \label{eq:si_word_length}
    \end{equation}
    
    In addition, when the margin is large enough, the small pull-down leakage from the `match' cells could change the search result as pointed out in the main text.
    This is because the conductance of the pull-down transistor is continuous with respect to the DL voltage ($G=f(V_\text{DL} ) $), and so the search range for one analog CAM cell in an array is affected by other cells attached to the same ML, described in \SIequation\ref{eq:si_word_range}.

    \begin{equation}
        G_\text{T} = f(V_\text{DL}) < G_\text{th} - (N-1)\cdot G_\text{T, OFF}
        \label{eq:si_word_range}
    \end{equation}
    where $G_\text{th}$ is the criteria to differentiate the `match' and `mismatch' case. It is clear that other cells attached to the same ML equivalently shift the criteria, and therefore the search range. The amount of the change is reflected by $\partial G_\text{T}/\partial V_\text{DL} $, \textit{i.e.} the conductance sensitivity to the change of analog voltage signal. The sensitivity can be written as $(\alpha S_{s})^{-1}$, where $\alpha$ is the ratio between changes in $V_\text{DL}$ and $V_\text{G}$, and $S_{s}$ is the subthreshold swing slope of the transistor. 
    As an example, considering a typical value for the $V_\text{DL}$ to $V_\text{G}$ ratio of 0.1 (simulated data shown in \Mfigure\ref{fig:simulation}d and \ref{fig:simulation}e), and $S_s$ of \SI{100}{\milli\volt}/dec, the equation gives the overall sensitivity of about dec/\SI{10}{\milli\volt}. 
    In simulation, the change in the search range simulated in \Mfigure\ref{fig:analysis} is tens of mV, which is fairly consistent with our analysis here. 
    Therefore, employing the volatile threshold switching memristor with small sub-threshold swing great improves the performance, as simulated in \Mfigure\ref{fig:emerging} and described in the main text. 
\end{sinote}

\begin{sinote}
    Analog CAM with emerging threshold switching memristor
    \label{not:si_emerging}

    Typically, sub-threshold current leakage through the ML pull-down transistors limits the maximum CAM word length and the number of stored bits per cell. To improve the maximum CAM work length in our analog CAM, we propose to replace the standard ML pull-down transistor with volatile threshold switching (TS) memristors\citesicite{s_midya2017am,s_kim2012nbo2,s_son2011vo2} which results in greatly reduced sub-threshold current leakage.
    Our analog CAM cell is converted to a circuit that is composed of four transistors, two non-volatile memristors, and two volatile TS memristors, as shown in \SIfigure\ref{fig:emerging}a.
    In contrast to our first implementation using pull-down transistors to discharge the ML for a mismatch result, a search operation in this case starts with ML at ground, and the ML is charged up only for mismatch cases. 
    The performance of the proposed cell is evaluated in simulation under 180 nm design rules which we used for the tapeout experiment. 
    The TS memristor is modelled with Verilog-A on data extracted from published experimental data\citesicite{s_midya2017am}. 
    Although TS memristors in the literature may suffer from endurance limits, the following analysis is aimed at providing a direction for even further performance improvements, with future work needed to fully explore these trade-offs.

    \SIfigure\ref{fig:emerging}b shows the simulated current-voltage (IV) curve for the TS memristor, from which one sees a significantly smaller sub-threshold swing than MOSFET transistors, thereby greatly decreasing the sub-threshold current leakage on the ML. 
    The ML voltage sensed at 50 ns after the search starts (\SIfigure\ref{fig:emerging}) shows a match for $V_\text{DL}$ between \SI{0.53}{\volt} and \SI{0.65}{\volt} (see \SIsection\ref{not:si_emerging} for additional details).
    As with the previous analog CAM cell, the simulated analog cell can be configured to match different $V_\text{DL}$ ranges by programming different memristor conductances. \SIfigure\ref{fig:emerging}d shows the successful analog CAM cell with 16 discrete programmable matching states.
    Further simulations (\SIfigure\ref{fig:emerging}e and f) of the analog CAM arrays show that the change in the matching voltage range moves by less than \SI{0.01}{\volt} with columns of up to 72, indicating the capability to store and search 5-6 bits of information, showing a significant improvement from conventional designs with pull-down transistors.
\end{sinote}

\begin{sinote}
    Comparison between the range search with TCAM and multibit analog CAM
    \label{not:si_range_cell}

    Here we consider the CAMs that are used in a network router for classifying a random range in a 16-bit Class B IP address space (0-65535). 
    In a TCAM, a continuous range can be represented by storing `X' in the least significant bits - for example, 01XX is a range between the binary number of 0100 and 0111 (or decimal number 4-7). 
    However, in most cases a random range will need to be split into multiple entries to fit in a TCAM. 
    Taking a random range between 385 and 58630 (or hexadecimal number 0181 - E506), we observe that this range can be implemented by a 21$\times$16 TCAM array as shown in \SIfigure\ref{fig:si_range_search}. 
    Although it may seem to be an acceptable overhead in this example here, this may not be the case when the ranges in a larger space are considered (e.g. 128 bit for IPv6). Here, we restrict ourselves to this smaller example for comparison purposes. 

    First, we observe that the whole TCAM array described above for a range search can be replaced by a single 16-bit analog cell. 
    With the present design, it is challenging to realize a 16-bit analog CAM cell currently, but we've shown that it is very feasible for a 3-bit or 4-bit cell to be implemented, as demonstrated in \Mfigure\ref{fig:simulation}e,f. 
    With limited bit-precision analog CAM cells, the range can be split in a similar way as in TCAM entries. 
    \SIfigure\ref{fig:si_range_search}b and \SIfigure\ref{fig:si_range_search}c show implementations with 3-bit analog CAM and 4-bit analog CAM respectively. 
    In the table, `X' is similar to that in the TCAM, which matches everything, i.e. 0-7 in a 3-bit cell or 0-15 in a 4-bit cell. 
    $\{n-m\}$ represent the cell is configured to match part of the range between $n$ and $m$.
    From the figure, one sees that by using a multi-bit analog CAM, both columns and rows can be compressed, leading to a reduction from 336 TCAM cells to 54 3-bit cells, 24 4-bit cells, or 6 8-bit cells. 
    In addition, there are only six transistors in an analog CAM cell, while 16 in a SRAM-based one. 
    With these factors taken into consideration, the overall transistor count required for this specific function results in a 37$\times$ reduction. 
    Assuming a TCAM implementation with \SI{0.70}{\square\micro\meter} per TCAM cell area overhead in a standard library under the same \SI{16}{\nano\meter} technology node, the proposed range search functionality occupies \SI{235.20}{\square\micro\meter} chip area while only \SI{12.48}{\square\micro\meter} with our analog CAM, leading to 18.8$\times$ reduction in area. 
    The decreased area reduction comparing to that in transistor count is due to the fact that the reference SRAM layout utilizes the foundry's SRAM-specific layout rules, while our present analog CAM layout follows a more conservative logic rule.
    We also expect a similar reduction in operational power with the analog CAM cell in comparison to conventional TCAMs, as a major portion of the dynamic energy consumption for a CAM operation is charging parasitic wire capacitances, and the reduced cell count and area also results in shorter wires and reduces total wire capacitance.

\end{sinote}

\begin{sinote}
    Energy estimation with a custom designed digital-to-analog converter
    \label{not:si_dac}

    To evaluate the search energy consumption, our analog CAM array simulations measure the consumed power during search operations by integrating current from all power supplies.
    As the evaluated IP routing application handles digital signals, digital-to-analog converters (DAC) are included in the analysis. The DACs impose additional overhead in terms of both chip area and energy, but the analysis below shows that the DAC overhead is not overwhelming.
    Importantly, for applications handling analog signals directly, digital-analog signal conversion is not required at all with our analog CAM, suggesting a promising application space. In contrast, analog-to-digital (ADC) converters would be required when using digital SRAM-based TCAMs, and we note ADCs are usually much more expensive (area/power) than DACs. 

    The DAC we use in our evaluation is a simple 4-bit current-steering design, the schematic of which is shown in \SIfigure\ref{fig:si_dac}.
    There are two major portions in the DAC design, with the first being the circuit mirrors input for $I_\text{REF}\times1$ and $I_\text{REF}\times 4$ respectively (shown in panel a). 
    The input current mirrors are shared globally across all the DAC channels, while the current mirror outputs are selectively turned on to convert the digital inputs that are applied on the switch transistors to analog outputs (panel b). 
    The simulated curves for the DAC operation are shown in \SIfigure\ref{fig:si_dac}c. 

    While unoptimized, the total energy consumption during search operations in an $86\times12$ analog CAM array is estimated to be 0.52 fJ per search per cell according to the simulations (\SItable\ref{tb:si_energy}), and will be smaller in practice due to larger arrays and their reduced average DAC peripheral cost. 
    If we convert the performance number in the aforementioned search table for IP routing, the same search function in the $21\times16$ TCAM table will consume 12.48 fJ per search in a $6\times4$ analog array, leading to a 0.037 fJ per search per equivalent TCAM bit. 
    Additional energy improvements are expected for applications that handle analog signals directly.

\end{sinote}

\begin{sinote}
    Decision Tree model mapped in our proposed analog CAM
    \label{not:decision_tree}

    Decision trees with binary and non-binary classification features can be implemented in the analog CAM directly by mapping each root to leaf path to a row in the analog CAM. Logically, each root-to-leaf path traverses a series of nodes with Boolean ANDs between elements in a given input feature vector (\SIfigure\ref{fig:si_dt}a). Since AND is commutative, we can reorder the nodes such that feature variables are processed in the same order for all paths\citesicite{s_buschjager2017decision}. Nodes for the same feature are combined into one node and ``don't care'' nodes can be inserted for features absent from a specific path, such that each path is of equal length. This representation can then be directly mapped to the analog CAM array, with each root to leaf path a row (see \SIfigure\ref{fig:si_dt}). As the matching row can directly drive the readout of the  classification result, tree traversal becomes a one-cycle operation (\SIfigure\ref{fig:si_dt}b). As each of the split outcomes of the tree are mutually exclusive, only one root-to-leaf path, or one row in the analog CAM, will ``match'' for a given feature vector. A collection of analog CAM arrays for decision trees with a local direct classification lookup can be used to implement ensemble tree-based models which are popular machine learning models.
    
\end{sinote}


\section{Supplementary References}
\bibliographystylesicite{naturemag}
\bibliographysicite{sicite}

\end{document}